\documentclass[twocolumn,showpacs,preprintnumbers,superscriptaddress,amsmath,floatfix,amssymb,secnumarabic]{revtex4}
\usepackage[colorlinks=true]{hyperref}
\usepackage{graphicx}
\graphicspath{ {./plots/} }

\newcommand{\lr}[1]{ \left( #1 \right) }
\newcommand{\lrs}[1]{ \left[ #1 \right] }
\newcommand{\lrc}[1]{ \left\{ #1 \right\} }
\newcommand{\vev}[1]{ \langle \, #1 \, \rangle }

\newcommand{\tr}{ {\rm Tr} \, }

\newcommand{\ket}[1]{ \, | #1 \rangle }
\newcommand{\bra}[1]{ \langle #1 | \, }

\newcommand{\diag}[1]{ {\rm diag} \, \left( #1 \right) }

\newcommand{\expa}[1]{ \exp{\left( #1 \right)} }

\newcommand{\mr}{\ensuremath{m_r}}           % Mass renormalisation
\newcommand{\mi}{\ensuremath{m_i}}           % Parity breaking mass
\newcommand{\mb}{\ensuremath{m^{(0)}}}        % Bare mass
\newcommand{\mua}{\ensuremath{\mu_A}}        % Chiral chemical potential
\newcommand{\mub}{\ensuremath{\mu_A^{(0)}}}   % Bare chiral chemical potential
\newcommand{\ls}{\ensuremath{L_s}}           % Linear lattice size
\newcommand{\mP}{\ensuremath{m_{\pi}}}        % Pion mass
\newcommand{\scme}{\ensuremath{\sigma_{CME}}} % sigma_cme
\newcommand{\smax}{\ensuremath{\sigma_{max}}} % sigma_max
\newcommand{\B}{\ensuremath{\operatorname{B}^0}}
\newcommand{\eV}{\ensuremath{\text{eV}}}
\newcommand{\meV}{\ensuremath{\operatorname{meV}}}

\begin{document}
\sloppy

\title{Chiral magnetic conductivity in an interacting lattice model of a parity-breaking Weyl semimetal}

\author{P.~V.~Buividovich}
\email{pavel.buividovich@physik.uni-regensburg.de}
\affiliation{Regensburg University, D-93053 Regensburg, Germany}

\author{M.~Puhr}
\email{matthias.puhr@physik.uni-regensburg.de}
\affiliation{Regensburg University, D-93053 Regensburg, Germany}

\author{S.~N.~Valgushev}
\email{semen.valgushev@physik.uni-regensburg.de}
\affiliation{Regensburg University, D-93053 Regensburg, Germany}

\date{September 22nd, 2015}
\begin{abstract}
 We report on the mean-field study of the Chiral Magnetic Effect (CME) in static magnetic fields within a simple model of a parity-breaking Weyl semimetal given by the lattice Wilson-Dirac Hamiltonian with constant chiral chemical potential. We consider both the mean-field renormalization of the model parameters and nontrivial corrections to the CME originating from re-summed ladder diagrams with arbitrary number of loops. We find that on-site repulsive interactions affect the chiral magnetic conductivity almost exclusively through the enhancement of the renormalized chiral chemical potential. Our results suggest that nontrivial corrections to the chiral magnetic conductivity due to inter-fermion interactions are not relevant in practice, since they only become important when the CME response is strongly suppressed by the large gap in the energy spectrum.
\end{abstract}

\maketitle

\section{Introduction}
\label{sec:intro}

 Anomaly-driven quantum transport phenomena in Dirac and Weyl semimetals, such as the Chiral Magnetic Effect (CME) \cite{Kharzeev:08:2} and the anomalous Hall effect \cite{Burkov:14:1}, are nowadays in the focus of active theoretical and experimental studies. On the one hand, these phenomena are very interesting from a theorist's point of view since they are nontrivial macroscopic manifestations of the universal laws of quantum physics at the microscopic level, much like the phenomena of superconductivity and superfluidity. On the other hand, implementation of anomalous transport in real materials would offer an intriguing possibility of a dissipation-free transport of spin-polarized current. An experimental signature of the Chiral Magnetic effect is the quadratic enhancement of the electric conductivity in the direction of the applied constant magnetic field, which was originally predicted in \cite{Nielsen:83:1} and has recently been observed in the Dirac semimetals Bi$_{1-x}$Sb$_x$ \cite{Kim:13:1}, ZrTe$_5$ \cite{Kharzeev:14:1} and Na$_3$Bi \cite{Xiong:15:1}. Very similar behavior of the electric conductivity in the direction of the magnetic field has also been found in holographic models \cite{Landsteiner:14:1,Landsteiner:15:1} and numerically in lattice QCD \cite{Buividovich:10:1}.
% \cite{Kharzeev:14:1} - 20 K temperature
% \cite{Kim:13:1} - 2 K
% \cite{Xiong:15:1} - around 5 K

 The Chiral Magnetic Effect is characterized by the chiral magnetic conductivity $\scme$, which relates the induced electric current $\vec{j}$ and the external magnetic field $\vec{B}$:
\begin{eqnarray}
\label{cmc_def}
 \vec{j} = \scme \, \vec{B} ,
\end{eqnarray}
where we assume the natural units for all the dimensionful quantities. For free Dirac fermions, the chiral magnetic conductivity is \cite{Kharzeev:08:2}
\begin{eqnarray}
\label{cmc_theory0}
 \scme = \frac{N_f \, \mu_A}{2 \pi^2} ,
\end{eqnarray}
where $N_f$ is the number of Dirac cones in the energy spectrum and $\mu_A$ is the chiral chemical potential which parameterizes the imbalance of densities of left- and right-handed fermions. In experimental studies of the magneto-resistivity of Dirac/Weyl semimetals, this chirality imbalance is also generated due to the quantum anomaly and within a simple relaxation time approximation can be estimated as \cite{Kharzeev:14:1}
\begin{eqnarray}
\label{muA_estimate}
 \mu_A = \frac{3 \alpha v_F^3 \tau}{4 \pi^2} \frac{\vec{E} \cdot \vec{B}}{T^2} ,
\end{eqnarray}
where $\vec{E}$ is the external electric field which probes the DC conductivity of the material, $v_F$ is the Fermi velocity, $\tau$ is the relaxation time of axial charge, and $T$ is the temperature. Combining this formula with (\ref{cmc_def}) and (\ref{cmc_theory0}), it is straightforward to see that the electric conductivity should grow as a square of the magnetic field.

 The relations (\ref{cmc_def}) and (\ref{cmc_theory0}) can be also reproduced in various models of parity-breaking Weyl semimetals \cite{Zyuzin:12:1,Goswami:12:1,Goswami:13:1,Landsteiner:13:1,Basar:13:1,Chang:14:1}. Strictly speaking, for lattice models this formula is only valid in the limit of a spatially uniform magnetic field which slowly changes in time \cite{Burkov:13:1} - in the case of a static spatially uniform magnetic field $\scme$ vanishes \cite{Vazifeh:13:1}, in accordance with the general requirements of the gauge invariance of the effective action \cite{Rubakov:10:1,Yamamoto:15:1}. However, if the static external magnetic field is allowed to be spatially non-uniform, $\scme$ becomes some non-trivial function of the wave vector $\vec{k}$ of the external field. Moreover, if the chiral symmetry is not spontaneously broken, the asymptotic value of $\scme\lr{\vec{k}}$ at large $\vec{k}$ is related to the axial anomaly coefficient in a nontrivial way \cite{Buividovich:13:8} and approaches the universal value (\ref{cmc_theory0}), but with a \textit{minus} sign. This change of sign as compared to the most commonly quoted result (\ref{cmc_theory0}) is related to the vector current conserving regularization of $\scme$ \cite{Ren:11:1,Buividovich:13:8} and does not affect the relation between $\scme$ and the anomaly coefficient.

 So far, most calculations of the chiral magnetic conductivity were performed without taking into account the inter-electron interactions. This approximation is not unreasonable, since in the hydrodynamics, kinetic theory or Fermi liquid approximations anomalous transport coefficients take universal values even in the presence of inter-fermion interactions \cite{Son:09:1,Sadofyev:10:1,Banerjee:12:1,Jensen:12:1,Stephanov:12:1}. However, in recent years it has been realized that the chiral magnetic conductivity (as well as the other anomalous transport coefficients) can still get nontrivial corrections due to inter-fermion interactions in two important situations: first, when the chiral symmetry is spontaneously broken \cite{Buividovich:13:8,Buividovich:14:1} and second, when the electric current is coupled to a dynamical gauge field \cite{Miransky:13:1,Jensen:13:1,Gursoy:14:1,Buividovich:14:1}. In the first case, the appearance of massless Goldstone modes violates the applicability of the hydrodynamical approximation, and the effective mass term which mixes chiralities makes the notion of a Fermi surface at finite $\mu_A$ ill-defined \cite{Buividovich:13:8}. In the second case, the correlators of electric current $\vec{j}$ receive perturbative QED corrections, which result in a renormalization of $\scme$.

 Clearly, both of these cases are relevant for real Weyl semimetals. Electrostatic inter-electron interactions in Dirac and Weyl semimetals are effectively enhanced by a factor of inverse Fermi velocity $v_F \sim 10^{-3} \ldots 10^{-2}$ \cite{Kim:13:1,Kharzeev:14:1}, thus the effective QED coupling constant $\alpha_{eff} \sim \alpha_{QED}/v_F \sim \frac{1}{137 \, v_F}$ is of order of one. Therefore in Weyl semimetals one typically deals with a strongly coupled fermionic gas which might become unstable towards spontaneous chiral symmetry breaking. On the other hand, inter-electron interactions in the context of condensed matter physics are always mediated by dynamical photon fields, therefore even interactions which are too weak to trigger spontaneous symmetry breaking can in general result in perturbative corrections to $\scme$.

 One should also mention that the classical dynamics of an electromagnetic field coupled to the plasma of chiral fermions at finite chiral chemical potential $\mu_A$ is likely to be unstable with respect to the decay of the chiral chemical potential at the expense of increasing magnetic helicity, as the studies within the frameworks of the chiral kinetic theory \cite{Yamamoto:13:1,Manuel:15:1}, magnetostatics \cite{Sadofyev:13:1,Sadofyev:13:2} and the AdS/CFT correspondence \cite{Kim:10:1,Zamaklar:11:1,Bayona:12:1} suggest. Therefore modelling the parity-breaking Weyl semimetals with a time-independent Dirac-like Hamiltonian with finite $\mu_A$ is a certain idealization valid at timescales much shorter than the typical decay time of chirality imbalance. The origin of this chiral plasma instability is the coupling to the magnetic field, which is suppressed as $1/v_F$ in Dirac and Weyl semimetals. Hence one can expect that the typical timescale of the decay of $\mu_A$ is rather long, much longer than the timescales at which e.g. the formation of the chiral condensate due to strong electrostatic interactions takes place. Recent real-time simulations of chiral plasma instability \cite{Buividovich:15:2} support this expectation. Another situation in which constant $\mu_A$ might be a reasonable approximation is when the decay of chirality imbalance due to chiral plasma instabilities or other chirality-changing processes is compensated by a constant inflow of chirality in parallel electric and magnetic fields \cite{Kharzeev:14:1,Parameswaran:13:1}. Since in experiments one typically deals with a strong constant magnetic field \cite{Kim:13:1,Kharzeev:14:1,Xiong:15:1} and an electric field of rather low frequency, a reasonably good approximation is to assume that the magnetic field and the electric current induced by it are also static. In this case, the calculation of the current becomes considerably simpler and can be carried out entirely in the equilibrium-state formalism.
Another motivation for considering the static limit is that at sufficiently small frequencies and momenta the difference between the static and time-dependent chiral magnetic conductivities vanishes once one takes into account relaxation processes \cite{Yee:14:1}.

 In this paper we study the effect of strong inter-electron interactions on the static chiral magnetic conductivity in a parity-breaking Weyl semimetal. In particular, we are interested in the question whether strong inter-electron interactions can lead to significant deviations of the chiral magnetic conductivity from the universal value (\ref{cmc_theory0}). A particularly interesting possibility is that the interactions can strongly enhance the CME response, as suggested by the mean-field analysis for the continuum Dirac fermions \cite{Buividovich:14:1}. Understanding the influence of inter-electron interactions on the chiral magnetic effect is thus important for the correct interpretation of the recent experiments \cite{Kim:13:1,Kharzeev:14:1,Xiong:15:1}.

  However, the continuum Dirac Hamiltonian is only the first, and rather uncontrollable, approximation for the realistic tight-binding models of Weyl semimetals. First of all, the introduction of a momentum cutoff for a continuum Dirac Hamiltonian violates the vector current conservation, which should then be manually restored by adding the so-called Bardeen counterterm \cite{Buividovich:14:1,Rubakov:10:1}. Second, the continuum Dirac Hamiltonian possesses exact chiral symmetry, which is typically violated in lattice systems at energies sufficiently higher than the energies of the Dirac points. This violation drastically changes the pattern of spontaneous symmetry breaking, in particular, turning the phase with spontaneously broken continuous $U\lr{1}$ chiral symmetry and massless Goldstone modes into the ``axionic insulator'' phase (or the Aoki phase in the terminology of lattice QCD) in which only discrete $Z_2$ $\mathcal{C}\mathcal{P}$ symmetry is spontaneously broken and no Goldstones emerge \cite{Aoki:84:1,Buividovich:14:2}. Therefore it is important to go beyond the Dirac cone approximation and to perform a consistent calculation of the chiral magnetic conductivity directly in the realistic tight-binding model of a Weyl semimetal.

In this paper we follow \cite{Fu:07:1,Sekine:13:1,Sekine:13:2,Zhang:09:1,Araki:13:2,Araki:13:3,Vazifeh:13:1,Hosur:13:1} and use the Wilson-Dirac Hamiltonian as the simplest realistic lattice model of Dirac semimetals and/or topological insulators, which are then tuned into a Weyl semimetal phase by adding the parity- or time-reversal-breaking terms. In our case, parity is broken by the chiral chemical potential $\mu_A$. We will use the mean-field approximation both to find the ground state of our model and to calculate the static chiral magnetic conductivity. Since the static chiral magnetic conductivity is the static linear response of the electric current to a small static external magnetic field, in the mean-field approximation it can be related to the linearized response of the static mean-field condensates to the static external gauge field. Having in mind that recent experiments \cite{Kim:13:1,Kharzeev:14:1,Xiong:15:1} are performed at very low temperatures ($\sim 10^1 \, {\rm K}$), we also mostly concentrate on the case of zero temperature, and discuss the effect of finite temperature (which in most cases acts against CME) only briefly.

 In contrast to the results obtained in \cite{Buividovich:14:1} for the continuum Dirac Hamiltonian, we find that the only source of enhancement of the chiral magnetic conductivity is the renormalization of the chiral chemical potential $\mua$, and it never exceeds the universal value $\scme = \frac{\mua \, N_f}{2 \pi^2}$ with the renormalized $\mua$. The universal value (\ref{cmc_theory0}) is only approached in the regions of the phase diagram where our model features massless Dirac cones in the energy spectrum. We find that in these regions the corrections due to inter-electron interactions (apart from those originating from the renormalization of $\mua$) are very small, not larger than a few tenth of percent. Corrections due to interactions only become important when the energy spectrum has a large gap, and reach maximum in the phase with spontaneously broken $\mathcal{CP}$ symmetry.

 On the other hand, the phase structure of our lattice model in the parameter space of the chiral chemical potential and interaction strength is more similar to the one obtained for the continuum Dirac Hamiltonian \cite{Buividovich:14:1}. In particular, we find that interactions effectively enhance the chiral chemical potential $\mu_A$. Nonzero chirality imbalance also shifts the boundaries of the Aoki phase towards weaker inter-electron interactions - much like the critical interaction strength for spontaneous breaking of chiral symmetry is decreased in the presence of $\mu_A$ \cite{Buividovich:14:1}. The only qualitative difference with the continuum case is that in the lattice model we do not observe any Cooper-type instability towards spontaneous breaking of chiral symmetry at small inter-electron interactions. The origin of this discrepancy is quite clear: for the Wilson-Dirac Hamiltonian, continuous chiral symmetry is already explicitly broken at high energies. Preliminary mean-field studies of the phase diagram of Wilson-Dirac fermions with chirality imbalance and on-site inter-electron interactions were already reported by two of the authors in \cite{Buividovich:14:2}.

 The outline of this paper is the following: in Section \ref{sec:mean_field_general}, we present a general framework for the mean-field calculations of the phase diagram and of the chiral magnetic conductivity in our lattice model. In Section \ref{sec:phase_diagram}, we study the mean-field phase diagram of the Wilson-Dirac Hamiltonian with chirality imbalance in the parameter space of chiral chemical potential, bare Dirac mass and inter-electron interaction potential. In Section \ref{sec:cmc}, we perform the mean-field calculation of the static chiral magnetic conductivity in different regions of the phase diagram. In Section \ref{sec:experiment} we comment on the feasibility of the experimental measurements of the static chiral magnetic conductivity. We summarize the results of our study in the concluding Section \ref{sec:conclusions}.

\section{Mean-field approximation for the Wilson-Dirac Hamiltonian with on-site inter-electron interactions}
\label{sec:mean_field_general}

 The starting point of our study is the Wilson-Dirac Hamiltonian with chiral chemical potential $\mub$ and on-site inter-electron interactions with repulsive potential $U > 0$:
\begin{eqnarray}
\label{hwdirac_manybody}
 \hat{H} = \sum\limits_{x,y} \hat{\psi}^{\dag}_x h^{\lr{0}}_{xy} \hat{\psi}_y  + U \sum\limits_x \lr{\hat{\psi}^{\dag}_x \hat{\psi}_x - 2}^2 ,
\end{eqnarray}
where $\hat{\psi}^{\dag}_x$ and $\hat{\psi}_y$ are the fermionic creation and annihilation operators at the lattice sites $x$ and $y$ and $h^{\lr{0}}_{xy}$ is the single-particle Wilson-Dirac Hamiltonian on the cubic three-dimensional lattice. In the absence of external fields, it has the following form in momentum space:
\begin{eqnarray}
\label{hwdirac_singlepart_momentum}
 h^{\lr{0}}\lr{\vec{k}} = \sum\limits_{i=1}^{3} \alpha_i v_F \sin\lr{k_i}
 +  \nonumber \\ +
 2 r \, \gamma_0 \sum\limits_{i=1}^{3} \sin^2\lr{k_i/2}
 + \gamma_0 m^{\lr{0}}
 + \gamma_5 \mub ,
\end{eqnarray}
where $v_F$ is the Fermi velocity, $r$ is the Wilson coefficient, $m^{\lr{0}}$ is the Dirac mass term,
\begin{eqnarray}
\label{dirac_matrices}
 \alpha_k = \diag{\sigma_k, -\sigma_k}, \quad \gamma_5 = \diag{I, -I},
 \nonumber \\
 \gamma_0 = \left(
  \begin{array}{cc}
   0 & I \\
   I & 0 \\
  \end{array}
 \right)
\end{eqnarray}
are the Dirac matrices, $\sigma_k$, $k = 1, 2, 3$ are the Pauli matrices and $I$ is the $2 \times 2$ identity matrix. In (\ref{hwdirac_singlepart_momentum}), we have expressed all dimensionful quantities in units of the inverse lattice spacing $a^{-1}$, which is hence just unity in what follows. We also use the units with $\hbar = c = 1$. The single-particle Hamiltonian (\ref{hwdirac_singlepart_momentum}) is the simplest model which mimics most important features of the more realistic tight-binding models of Dirac and Weyl semimetals \cite{Fu:07:1,Sekine:13:1,Sekine:13:2,Zhang:09:1,Araki:13:2,Araki:13:3,Vazifeh:13:1,Hosur:13:1}, such as the emergent chiral symmetry at small energies.

 The explicit coordinate-space form of $h^{0}_{x,y}$ in (\ref{hwdirac_manybody}) with external gauge fields $\vec{A}\lr{\vec{x}}$ can be found by the Peierls substitution of the link phases $A_{x,k} = \int\limits_{x}^{x + \hat{k}} dz A_k\lr{z}$ into the finite-difference operators on the lattice:
\begin{eqnarray}
\label{hwdirac_singlepart}
 h^{0}_{x, y}
 =
  \sum\limits_{k=1}^{3} -i \alpha_k v_F \nabla_{k \, x y}
 +
   \frac{r \, \gamma_0}{2} \Delta_{x y}
 + \nonumber \\ +
   \gamma_0 m^{\lr{0}}
 + \gamma_5 \mub ,
\end{eqnarray}
where
\begin{eqnarray}
\label{lattice_nabla_delta}
 \nabla_{k \, x y}
 =
 \frac{1}{2} \lr{e^{i A_{x,k}} \delta_{x+\hat{k},y} - e^{-i A_{x-\hat{k},k}}\delta_{x - \hat{k},y}} ,
\nonumber \\
 \Delta_{x y} = \sum\limits_{k=1}^{3} \lr{2 - e^{i A_{x,k}} \delta_{x+\hat{k},y} - e^{-i  A_{x-\hat{k},k}}\delta_{x - \hat{k},y}}
\end{eqnarray}
are the lattice discretizations of the covariant derivative operator $\nabla_{k}$ and (minus) the Laplacian operator $- \nabla_k^2$ and $\hat{k}$ is a unit lattice vector in the direction $k$.

 In (\ref{hwdirac_manybody}) we include only instantaneous on-site inter-electron interactions. On the one hand, mean-field phase diagrams obtained with only on-site interactions in most cases qualitatively reproduce most important features of the phase diagrams with more realistic interaction potentials. On the other hand, the Hubbard-Stratonovich transformation which is an important ingredient of the mean-field analysis greatly simplifies for on-site interactions. We also neglect the magnetic part of electromagnetic interactions, which is suppressed as $1/v_F$, and treat the vector gauge potential $A_k\lr{x}$ in the Hamiltonian (\ref{hwdirac_manybody}) as a non-dynamical external field which only serves as a source for the static electric current. The instantaneous approximation for our interaction potential can also be justified by the smallness of the Fermi velocity $v_F$ in realistic systems.

 Moreover, as we show in Appendix \ref{apdx:fermi_velocity}, if we only consider instantaneous inter-fermion interactions, the dependence on the Fermi velocity amounts to simple linear scaling of the model parameters and observables. Therefore we set $v_F = 1$ and summarize the most important formulas from which the full dependence on $v_F$ can be restored in all our results in Appendix \ref{apdx:fermi_velocity}. The dependence on the Wilson coefficient $r$ in (\ref{hwdirac_singlepart}), (\ref{hwdirac_singlepart_momentum}) is strictly speaking mathematically nontrivial. However, in practice changing $r$ results only in a rescaling of the phase diagram in the direction of the bare Dirac mass $m^{\lr{0}}$ \cite{Aoki:84:1}, and no qualitative changes of the phase structure occur. For this reason we also set $r = v_F = 1$ in order to reduce the number of free parameters in our model and simplify the analysis.

 Below we summarize the basic steps leading to the mean-field approximation for the partition function with the Hamiltonian (\ref{hwdirac_manybody}). A more detailed derivation can be found in \cite{Buividovich:14:1}. In order to calculate the partition function $\mathcal{Z} = \tr \expa{-\hat{H}/T}$, we perform the standard Suzuki-Trotter decomposition of the exponent $\expa{-\hat{H}/T}$, splitting the Euclidean time $\tau \in \lrs{0 \ldots T^{-1}}$ into infinitesimal intervals of size $\Delta \tau$:
\begin{eqnarray}
\label{SuzukiTrotter}
 \tr\expa{-\hat{H}/T}
 = \nonumber \\ =
 \lim\limits_{\Delta\tau \rightarrow 0} \tr\lr{
 e^{-\Delta \tau \hat{H}_0}
 e^{-\Delta \tau \hat{H}_I}
 e^{-\Delta \tau \hat{H}_0}
 e^{-\Delta \tau \hat{H}_I} \ldots
 } ,
\end{eqnarray}
where $\hat{H}_0$ is the free part of the Hamiltonian (\ref{hwdirac_manybody}) and $\hat{H}_I  = U \sum\limits_x \lr{\hat{\psi}^{\dag}_x \hat{\psi}_x - 2}^2$ is the interaction term. Next we use the Hubbard-Stratonovich transformation to rewrite the exponents involving the interaction term as
\begin{eqnarray}
\label{HubbardStratonovich1}
 \expa{- \Delta \tau \, \hat{H}_I}
 = %\nonumber \\ =
 \int \prod\limits_x d\Phi_{x}\lr{\tau}
 \nonumber \\
 \expa{ - \Delta \tau \sum\limits_x \lr{
 \frac{\tr{\Phi_{x}^2\lr{\tau}}}{4 U}
 + \hat{\psi}^{\dag}_x \Phi_{x}\lr{\tau} \hat{\psi}_{x} } }  ,
\end{eqnarray}
where $\Phi_x\lr{\tau}$ is a Hermitian traceless matrix with two Dirac spinor indices \footnote{The vanishing of the trace of $\Phi_x$ is related to the non-renormalization of the chemical potential \cite{Buividovich:14:1}} and we include some trivial normalization factors into the definition of the integration measure $d\Phi_x\lr{\tau}$. In the following it will also be convenient to represent $\Phi_x$ as a sum over a basis set of $15$ traceless Hermitian spinor matrices:
\begin{eqnarray}
\label{Phi_basis_decomposition}
 \Phi_x = \sum\limits_{A = 1}^{15} \Gamma_A \Phi_{x,A},
 \nonumber \\
 \Gamma_A = \lrc{\gamma_5 \alpha_k, \gamma_0, \gamma_0 \gamma_5 \alpha_k, -i \gamma_5 \gamma_0, i \gamma_0 \alpha_k, \gamma_5, \alpha_k },
\end{eqnarray}
where $k = 1, 2, 3$ labels spatial directions. The matrices $\Gamma_A$ are normalized as $\tr\lr{\Gamma_A \Gamma_B} = 4 \delta_{AB}$, so that the action of the Hubbard-Stratonovich field reads $\tr{\Phi_{x}^2}/\lr{4 U} = \Phi_{x,A}^2/U$.

 After inserting the transformation (\ref{HubbardStratonovich1}) into the Suzuki-Trotter decomposition (\ref{SuzukiTrotter}), we can represent the partition function $\mathcal{Z} = \tr \expa{-\hat{H}/T}$ in terms of the partition function $\mathcal{Z}\lrs{\Phi_x\lr{\tau}}$ of a free fermion gas in the background of the space- and time-dependent field $\Phi_x\lr{\tau}$:
\begin{eqnarray}
\label{HubbardStratonovich2}
 \mathcal{Z} = \int\mathcal{D}\Phi_x\lr{\tau} \, \mathcal{Z}\lrs{\Phi_x\lr{\tau}}
 \times \nonumber \\ \times
 \expa{- \int\limits_{0}^{T^{-1}} d\tau \frac{\tr{\Phi_x^2\lr{\tau}}}{4 U} } \,
 ,
 \nonumber \\
 \mathcal{Z}\lrs{\Phi_x\lr{\tau}}
 = \nonumber \\ =
 \tr \mathcal{T}\expa{- \int\limits_{0}^{T^{-1}} d\tau \lr{\hat{H}_0 + \sum\limits_x \hat{\psi}^{\dag}_x \Phi_x\lr{\tau} \hat{\psi}_x} },
\end{eqnarray}
with the time ordering operator $\mathcal{T}$.

We now apply the mean-field approximation and replace the integral over $\Phi_x\lr{\tau}$ by its value $\Phi^{\star}_x\lr{\tau}$ at the saddle point of the path integral (\ref{HubbardStratonovich2}). Moreover, we assume that $\Phi^{\star}_x\lr{\tau} = \Phi^{\star}_x$ does not depend on the Euclidean time $\tau$. The free energy $\mathcal{F} = -T \ln \mathcal{Z}$ which corresponds to the partition function (\ref{HubbardStratonovich2}) is then simply the functional
\begin{eqnarray}
\label{free_energy_mf1}
 \mathcal{F}\lrs{\Phi_x} = \mathcal{F}_0\lrs{\Phi_x} + \sum\limits_x \frac{\tr{\Phi_x^2}}{4 U} ,
\end{eqnarray}
where one should take the value $\Phi_x = \Phi^{\star}_x$ which minimizes $\mathcal{F}\lrs{\Phi_x}$ and $\mathcal{F}_0\lrs{\Phi_x}$ is the free energy of a free fermion gas with an effective single-particle Hamiltonian
\begin{eqnarray}
\label{effective_sp_hamiltonian}
 h_{x y} = h^{\lr{0}}_{x y} + \Phi_x \delta_{x y} .
\end{eqnarray}
The explicit expression for $\mathcal{F}_0\lrs{\Phi_x}$ reads
\begin{eqnarray}
\label{free_energy_fermion_gas}
 \mathcal{F}_0\lrs{\Phi_x} = -T \sum\limits_i \ln\lr{1 + e^{-\epsilon_i/T}} ,
\end{eqnarray}
where $\epsilon_i$ are the energy levels of the single-particle Hamiltonian (\ref{effective_sp_hamiltonian}).

 The static electric current which enters the definition (\ref{cmc_theory0}) of the chiral magnetic conductivity can be expressed as the variation of the free energy $\mathcal{F}$ over the gauge potential $A_k\lr{x}$. More precisely, in the lattice model (\ref{hwdirac_manybody}) the operator of electric current density is associated with lattice links, and its expectation value is related to the derivative of the partition function over the link phase factors $A_{x,k}$: $\vev{j_{x,k}} = \frac{\delta \mathcal{F}}{\delta A_{x, k}}$. Obviously, this current should vanish in the absence of external gauge fields. The linearized response of the electric current to a small external gauge field is then given by the second variation of the free energy with respect to the vector gauge potential at zero external field:
\begin{eqnarray}
\label{current_linear_response1}
 \vev{ j_{x,k} } = \sum\limits_{y, l} \left.\frac{\delta^2 \mathcal{F}\lrs{A_{x, k}}}{\delta A_{x, k} \, \delta A_{y, l}}\right|_{A = 0} A_{y,l} ,
\end{eqnarray}
where we have explicitly restored the dependence of the free energy on the link phases $A_{x, k}$. After expressing the static external magnetic field as $B_i\lr{x} = \epsilon_{ijk} \frac{\partial}{\partial x_j} A_k\lr{x}$ and rewriting the relation (\ref{current_linear_response1}) in the momentum space, we arrive at the following Kubo relation for the static chiral magnetic conductivity (\ref{cmc_theory0}) as a function of the wave vector $k$ of an external magnetic field
\cite{Gynther:10:1, Landsteiner:11:2, Landsteiner:12:1}:
\begin{eqnarray}
\label{cmc_Kubo}
 \scme\lr{k_3}
  =  -\frac{i}{k_3} \, \frac{1}{L^3} \sum\limits_{x, y} e^{i k_3 \lr{x_3 - y_3}} \vev{ j_{x,1} \, j_{y,2} }
  = \nonumber \\  =
  \left.
 -\frac{i}{k_3} \, \frac{1}{L^3} \sum\limits_{x, y} e^{i k_3 \lr{x_3 - y_3}}
 \frac{\delta^2 \mathcal{F}\lrs{A_{x, k}} }{\delta A_{x,1} \, \delta A_{y,2}} \,
  \right|_{A_{x,i} = 0}   & &
\end{eqnarray}
For definiteness, we have assumed in the above expression that the wave vector of the external magnetic field is parallel to the third coordinate axis and the vector gauge potential is parallel to the second axis. In this case the magnetic field and the CME current are parallel to the first coordinate axis.

 In the mean-field approximation, we replace the full free energy $\mathcal{F} = -T \ln \mathcal{Z}$ by the minimal value of the mean-field free energy (\ref{free_energy_mf1}). When using (\ref{cmc_Kubo}) to calculate the chiral magnetic conductivity, one should now keep in mind that the extremum value of the Hubbard field $\Phi^{\star}_x \equiv \Phi^{\star}_x\lrs{A_{x,k}}$ which minimizes the free energy $\mathcal{F}\lrs{\Phi_x, A_{x, k}}$ given by (\ref{free_energy_mf1}) in general depends on the external gauge field $A_{x, k}$. The variation $\frac{\delta \Phi^{\star}_x\lrs{A_{x,k}}}{\delta A_{x,k}}$ of the extremum value $\Phi^{\star}_x\lrs{A_{x,k}}$ of the Hubbard-Stratonovich field over the external gauge field $A_{x,k}$ can be found as the derivative of an implicit function using the extremum equation
\begin{eqnarray}
\label{mf_extremum_eq}
 \left. \frac{\partial \mathcal{F}\lrs{\Phi_x, A_{x,k}}}{\partial \Phi_{x,A}} \right|_{\Phi^{\star}}  = 0 .
\end{eqnarray}
After some algebraic manipulations which are summarized in detail in \cite{Buividovich:14:1}, we find the following general expression for the second derivative over the vector gauge potential which enters (\ref{cmc_Kubo}):
\begin{eqnarray}
\label{mf_response}
 \frac{\delta^2 \mathcal{F}\lrs{\Phi^{\star}_x, A_{x, k}}}{\delta A_{x,i} \, \delta A_{y,j}}
 =
 \frac{\partial^2 \mathcal{F}\lrs{\Phi^{\star}_x, A_{x, k}}}{\partial A_{x,i} \, \partial A_{y,j}}
 - \nonumber \\ -
 \left. \sum\limits_{z,A,t,B} G_{z,A;t,B}
 \frac{\partial^2 \mathcal{F}\lrs{\Phi_x, A_{x, k}}}{\partial A_{x,i} \, \partial \Phi_{z,A}}
 \frac{\partial^2 \mathcal{F}\lrs{\Phi_x, A_{x, k}}}{\partial A_{y,j} \, \partial \Phi_{t,B}}
 \right|_{\Phi^{\star}_x} ,
\end{eqnarray}
where $G_{z,A;t,B}$ is the propagator of the Hubbard-Stratonovich field defined as:
\begin{eqnarray}
\label{HubbardPropagator}
 \sum\limits_{y,B} G_{x,A;y,B} \frac{\partial^2 \mathcal{F}\lrs{\Phi_x, A_{x, k}}}{\partial \Phi_{y,B} \partial \Phi_{z,C}}
 =
 \delta_{xz} \delta_{AC} .
\end{eqnarray}
Now the derivatives $\partial/\partial A_{x,i}$ and $\partial/\partial \Phi_{x,A}$ in (\ref{mf_response}) and (\ref{HubbardPropagator}) should be calculated as partial derivatives, that is, in the process of differentiation the mean-field free energy functional (\ref{free_energy_mf1}) should be considered as a functional of the two independent fields $A_{x,i}$ and $\Phi_{x,A}$.

 The first term in (\ref{mf_response}) describes the electromagnetic response of the non-interacting system with renormalized single-particle Hamiltonian (\ref{effective_sp_hamiltonian}). Since the parameters of the renormalized, rather than the bare, single-particle Hamiltonian are physically observable, this term does not really describe the effect of interactions. Only the second term in (\ref{mf_response}) describes nontrivial corrections to the current-current correlators due to inter-electron interactions. In the weak-coupling regime, it can be represented as an infinite sum of ladder diagrams \cite{Buividovich:14:1}.

 While the differentiation of the second term in (\ref{free_energy_mf1}) is trivial, the calculation of the second derivatives $\frac{\partial^2 \mathcal{F}_0}{\partial A_{x,i} \, \partial A_{y,j}}$, $\frac{\partial^2 \mathcal{F}_0}{\partial A_{x,i} \, \partial \Phi_{y,A}}$ and $\frac{\partial^2 \mathcal{F}_0}{\partial \Phi_{x,A} \, \partial \Phi_{y,B}}$ of the fermionic free energy (\ref{free_energy_fermion_gas}) is somewhat lengthy, and we summarize it in Appendix \ref{apdx:ffe_derivatives}. The basic idea of this calculation is that in the limit of static Hubbard-Stratonovich and vector gauge fields the derivatives of the single-particle eigen-energies $\epsilon_i$ which enter (\ref{free_energy_fermion_gas}) can be obtained simply using the second-order quantum mechanical perturbation theory for the effective single-particle Hamiltonian (\ref{effective_sp_hamiltonian}):
\begin{eqnarray}
\label{qm_pt}
 \frac{\partial \epsilon_i}{\partial \theta}
 =
 \bra{\Psi_i} \frac{\partial h}{\partial \theta} \ket{\Psi_i},
 \nonumber \\
 \frac{\partial^2 \epsilon_i}{\partial \theta \, \partial \xi}
 =
 \bra{\Psi_i} \frac{\partial^2 h}{\partial \xi \, \partial \theta} \ket{\Psi_i}
 + \nonumber \\ +
  \sum\limits_{j \neq i}
 \frac{\bra{\Psi_i} \frac{\partial h}{\partial \theta} \ket{\Psi_j} \bra{\Psi_j} \frac{\partial h}{\partial \xi} \ket{\Psi_i}}{\epsilon_i - \epsilon_j}
 + \nonumber \\ +
 \sum\limits_{j \neq i}
 \frac{\bra{\Psi_i} \frac{\partial h}{\partial \xi} \ket{\Psi_j} \bra{\Psi_j} \frac{\partial h}{\partial \theta} \ket{\Psi_i}}{\epsilon_i - \epsilon_j}
 \end{eqnarray}
where the variables $\theta$, $\xi$ can be either the link factors $A_{x,i}$ or the Hubbard-Stratonovich fields $\Phi_{x,A}$ and $\ket{\Psi_i}$ denotes the eigenstate of the single-particle Hamiltonian (\ref{hwdirac_singlepart}) which corresponds to the energy level $\epsilon_i$. Note also that since the link factors $A_{x,i}$ enter the single-particle Hamiltonian (\ref{hwdirac_singlepart}) in a nonlinear way, we should in general keep the terms involving its second derivative $\frac{\partial^2 h}{\partial \xi \, \partial \theta}$. These terms are absent for the continuum Dirac Hamiltonian \cite{Buividovich:14:1}, where the gauge vector potential enters linearly, essentially in the same way as the component of the Hubbard-Stratonovich field which corresponds to $\Gamma_A = \alpha_i$. It is precisely this difference which renders the current-current correlator in (\ref{current_linear_response1}) manifestly gauge-invariant and effectively replaces the subtraction of the Bardeen counterterm.

 Differentiating now the free energy of the free fermion gas (\ref{free_energy_fermion_gas}) and using the above expressions, we obtain
\begin{eqnarray}
\label{ffe_2nd_der_general}
 \frac{\partial^2 \mathcal{F}_0}{\partial \theta \, \partial \xi}
 =
 \sum\limits_i\lr{ \frac{\partial^2 f\lr{\epsilon_i}}{\partial \epsilon^2} \, \frac{\partial \epsilon_i}{\partial \theta} \, \frac{\partial \epsilon_i}{\partial \xi}
 +
  \frac{\partial f\lr{\epsilon_i}}{\partial \epsilon} \, \frac{\partial^2 \epsilon_i}{\partial \theta \, \partial \xi}}
 = \nonumber \\ =
 \sum\limits_i \frac{-1}{4 T \, \cosh^2\lr{\frac{\epsilon_i}{2 T}}}
 \bra{\Psi_i} \frac{\partial h}{\partial \theta} \ket{\Psi_i}
 \bra{\Psi_i} \frac{\partial h}{\partial \xi}    \ket{\Psi_i}
 + \nonumber \\ +
 \sum\limits_i n\lr{\epsilon_i} \bra{\Psi_i} \frac{\partial^2 h}{\partial \xi \, \partial \theta} \ket{\Psi_i}
 + \nonumber \\ +
 \sum\limits_{i \neq j} n\lr{\epsilon_i}
 \frac{\bra{\Psi_i} \frac{\partial h}{\partial \theta} \ket{\Psi_j} \bra{\Psi_j} \frac{\partial h}{\partial \xi} \ket{\Psi_i}}{\epsilon_i - \epsilon_j}
 + \nonumber \\ +
 \sum\limits_{i \neq j} n\lr{\epsilon_i}
 \frac{\bra{\Psi_i} \frac{\partial h}{\partial \xi}    \ket{\Psi_j} \bra{\Psi_j} \frac{\partial h}{\partial \theta} \ket{\Psi_i}}{\epsilon_i - \epsilon_j}
 ,
\end{eqnarray}
where we have denoted $f\lr{\epsilon} = -T \ln\lr{1 + e^{-\epsilon/T}}$ and $n\lr{\epsilon} = \partial f\lr{\epsilon}/\partial \epsilon = \frac{1}{e^{\epsilon/T} + 1}$ is the Fermi factor.

 Since in (\ref{cmc_Kubo}) we are calculating $\scme$ in the linear response regime (\ref{ffe_2nd_der_general}) in the absence of background magnetic fields, in the expression (\ref{ffe_2nd_der_general}) we should set $A_{x,i}$ to zero and the Hubbard-Stratonovich field $\Phi_x$ - to its mean-field value. In this paper, we assume that the mean-field configuration of the Hubbard-Stratonovich field is spatially homogeneous (in the following we discuss the validity of this assumption in more details). In this case, the eigenstates $\ket{\Psi_i}$ can be found explicitly, which reduces the calculation of the second derivatives (\ref{ffe_2nd_der_general}) which enter (\ref{mf_response}) to a simple numerical summation over lattice momenta. In Section \ref{sec:phase_diagram} we present the mean-field phase diagram of our model (\ref{hwdirac_manybody}). In Section \ref{sec:cmc} we then use the mean-field single-particle Hamiltonian to calculate the chiral magnetic conductivity (\ref{cmc_Kubo}) using the expressions (\ref{mf_response}) and (\ref{ffe_2nd_der_general}).

\section{Mean-field phase diagram in the presence of chirality imbalance}
\label{sec:phase_diagram}

 To compute the mean-field phase diagram of the model (\ref{hwdirac_manybody}) we assume that the saddle-point Hubbard-Stratonovich field $\Phi_x^{\star} = \Phi^{\star}$ is homogeneous and does not break rotational symmetry. Under this assumption the saddle-point values are restricted to the form
\begin{eqnarray}
  \label{restr_saddle_point}
  \Phi^{\star} = \lr{\mr-\mb}\gamma_0 + i \mi\gamma_0\gamma_5+\lr{\mua-\mub}\gamma_5,
\end{eqnarray}
where we introduced the renormalized mass $\mr$, the renormalized chiral chemical potential $\mua$ and the $\mathcal{C}\mathcal{P}$-breaking mass term $\mi$ which corresponds to the ``axion'' condensate $\vev{\hat{\psi}^\dagger \gamma_0 \gamma_5 \hat{\psi}}$. With the mean-field value (\ref{restr_saddle_point}) for the Hubbard-Stratonovich field and in the limit of vanishing temperature the free energy density becomes
\begin{eqnarray}
  \label{eq:mean_field_free_energy}
  \frac{\mathcal{F}\lrs{\Phi^{\star}}}{\ls^3} = \frac{1}{\ls^3}\sum\limits_{\varepsilon < 0} \varepsilon
  + \nonumber \\ +
  \frac{(\mb-\mr)^2 + \mi^2 +(\mub-\mua)^2}{U}.
\end{eqnarray}
The first term is a sum  over all negative energy levels
\begin{eqnarray}
\label{eq:energy_levels}
 \varepsilon = \varepsilon_{s,\sigma}\lr{\vec{k}} = s\sqrt{\lr{S - \sigma\mua}^2 + \mi^2 + \lr{\mr+W}^2}
\end{eqnarray}
of the effective single particle Hamiltonian (\ref{effective_sp_hamiltonian})  and for the sake of a compact notation we define $s = \pm 1$,$\sigma = \pm 1$,  $S=\sqrt{\sum\limits_{k_i}\sin^2(k_i)}$ and $W=\sum\limits_{k_i}2r\sin^2(k_i/2)$.

 Before discussing the phase diagram of our model, it is useful to briefly recollect some basic facts about the Wilson-Dirac Hamiltonian (\ref{hwdirac_singlepart_momentum}). In the context of lattice QCD Wilson added the term proportional to $\Delta_{xy}$ to get rid of the so-called ``doublers'', unwanted fermionic degrees of freedom that should not be present in the continuum limit of lattice QCD. By adding the Wilson term to the naive discretization the doublers acquire a mass that depends on the inverse lattice spacing. In the continuum limit the doublers become infinitely heavy and decouple from the theory. We use Wilson-Dirac fermions to model the band structure in a crystal and use a fixed lattice spacing that was set to one in equation (\ref{lattice_nabla_delta}). In this context, the Wilson term $\Delta_{xy}$ reflects the fact that in Dirac or Weyl semimetals the continuum chiral symmetry is only an emergent symmetry at low energies, which is explicitly broken away from the Dirac/Weyl points. The mass of the doubler fermions is given by  the poles of the propagator in momentum space, which are located at positions where $\sum_{k_i} \sin{(k_i)}^2= 0$. Since the momenta lie in the Brillouin zone, i.e. $k_i \in (-\pi,\pi]$, the mass of the doublers reads
\begin{eqnarray}
  \label{eq:doubler_mass}
  m_d = \mr + 2 \lambda,
\end{eqnarray}
where $\lambda$ is the number of momentum components $k_i$ with $k_i = \pi$ and we assume that the Wilson parameter ${r=1}$ according to the discussion in the previous Section. In three dimensions $\lambda$ can take the values \mbox{$\lambda = 0, 1, 2, 3$}. There is only one way to achieve $\lambda = 0$ and $\lambda =3$, all entries of $\vec{k}$ have to be the same. For $\lambda = 1$ we have three possibilities to place the entry $\pi$ in a three dimensional vector,
\begin{eqnarray}
  \label{eq:doubler_k}
  \vec{k} = \begin{pmatrix}
              0,0,\pi
            \end{pmatrix}^T,
  \vec{k} = \begin{pmatrix}
               0,\pi,0
             \end{pmatrix}^T
  \text{ and }
  \vec{k} = \begin{pmatrix}
               \pi,0,0
             \end{pmatrix}^T.
\end{eqnarray}
Analogous for $\lambda = 2$ there are three ways to place the $0$. This means that for a bare mass of $\mr = 0$ and $\mr = -6$ we have $N_f = 1$ Dirac cones in the energy spectrum of our model, whereas there are $N_f = 3$ Dirac cones for $\mr = -2$ and $\mr = -4$. It follows from equation (\ref{eq:doubler_mass}) that non-trivial phases with massless excitations only exist for  \mbox{$-6 \leq \mb \leq 0$} and we restrict our study to this parameter range.

 After setting $r = v_F = 1$ and expressing all dimensionful quantities in units of the lattice spacing we are thus left with three free parameters in our model: the bare mass $\mb$, the inter-fermion interaction potential $U$ and the bare chiral chemical potential $\mub$. To compute the phase diagram we scan over these parameters and find the renormalized values of the parameters $\mr$, $\mi$ and $\mua$ of our model by numerically minimizing the free energy (\ref{eq:mean_field_free_energy}) with respect to them.

\begin{figure}[h!tpb]
  \centering
  \includegraphics[width=0.9\linewidth]{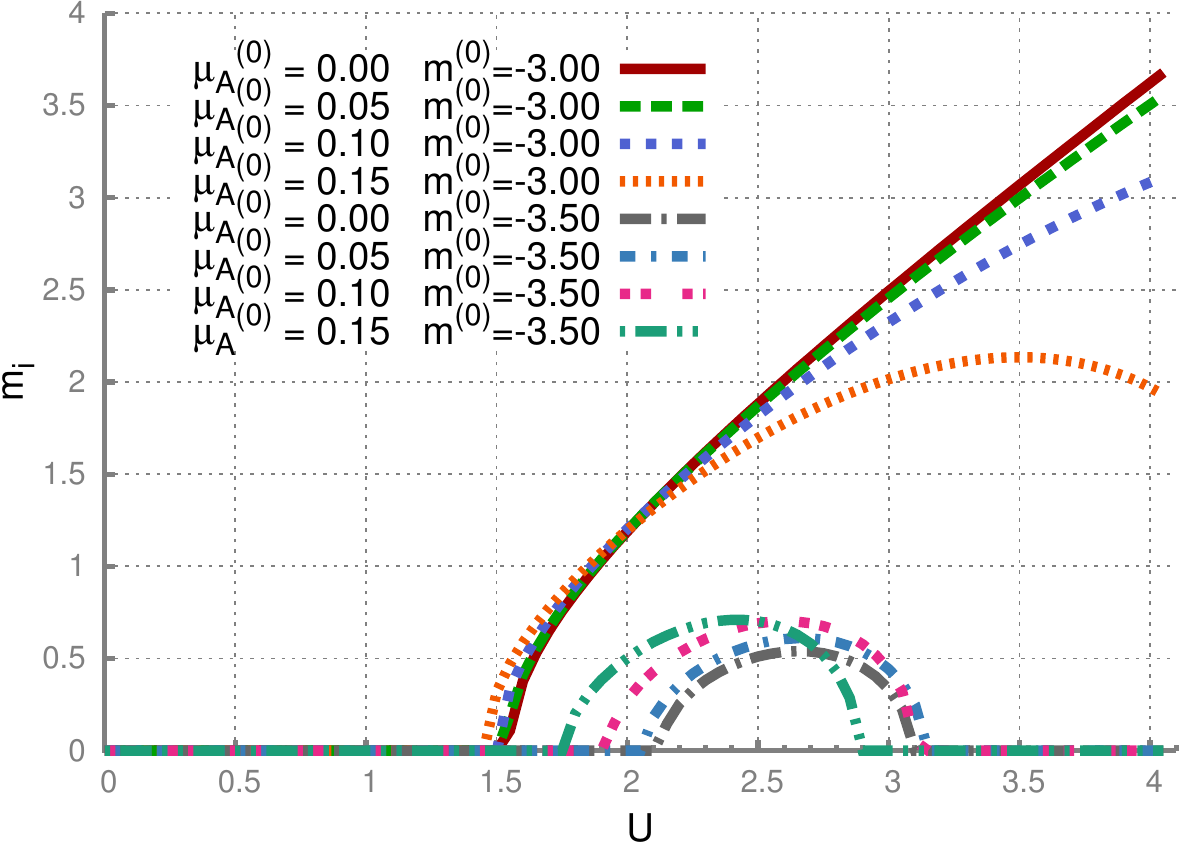}
  \caption{The $\mathcal{C}\mathcal{P}$-breaking mass term $\mi$ in (\ref{restr_saddle_point}) as a function of interaction potential $U$ for different values of the bare chiral chemical potential $\mub$.}
  \label{fig:mi}
\end{figure}

 Our model exhibits two distinct phases: A $\mathcal{CP}$-symmetric phase with $\mi = 0$ and the ``Aoki phase'' where the $\mathcal{C}\mathcal{P}$ symmetry is spontaneously broken by a nonzero $\mathcal{C}\mathcal{P}$-breaking effective mass term $m_i \neq 0$. In our lattice model, this spontaneous breaking of a discrete $\mathcal{C}\mathcal{P}$ symmetry is a remainder of the spontaneous breaking of continuous chiral symmetry which is now only an emergent low-energy symmetry. In the context of condensed matter physics, the Aoki phase is the ``axionic insulator'' phase with a nonzero condensate of an effective axion field (see e.g. \cite{Maciejko:13:1, Li:09:1,Sekine:14:1} for some examples of effective axion fields in condensed matter systems). On Fig.~\ref{fig:mi} we show the mean-field value of the $\mathcal{C}\mathcal{P}$-breaking mass term $m_i$ as a function of interaction potential $U$ for different values of the bare chiral chemical potential $\mub$. For all values of the chiral chemical potential $\mub$ the transition between the normal insulator/semimetal phase and the Aoki phase with $m_i \neq 0$ appears to be a sharp second-order phase transition.

\begin{figure}[h!tpb]
  \centering
  \includegraphics[width=0.9\linewidth]{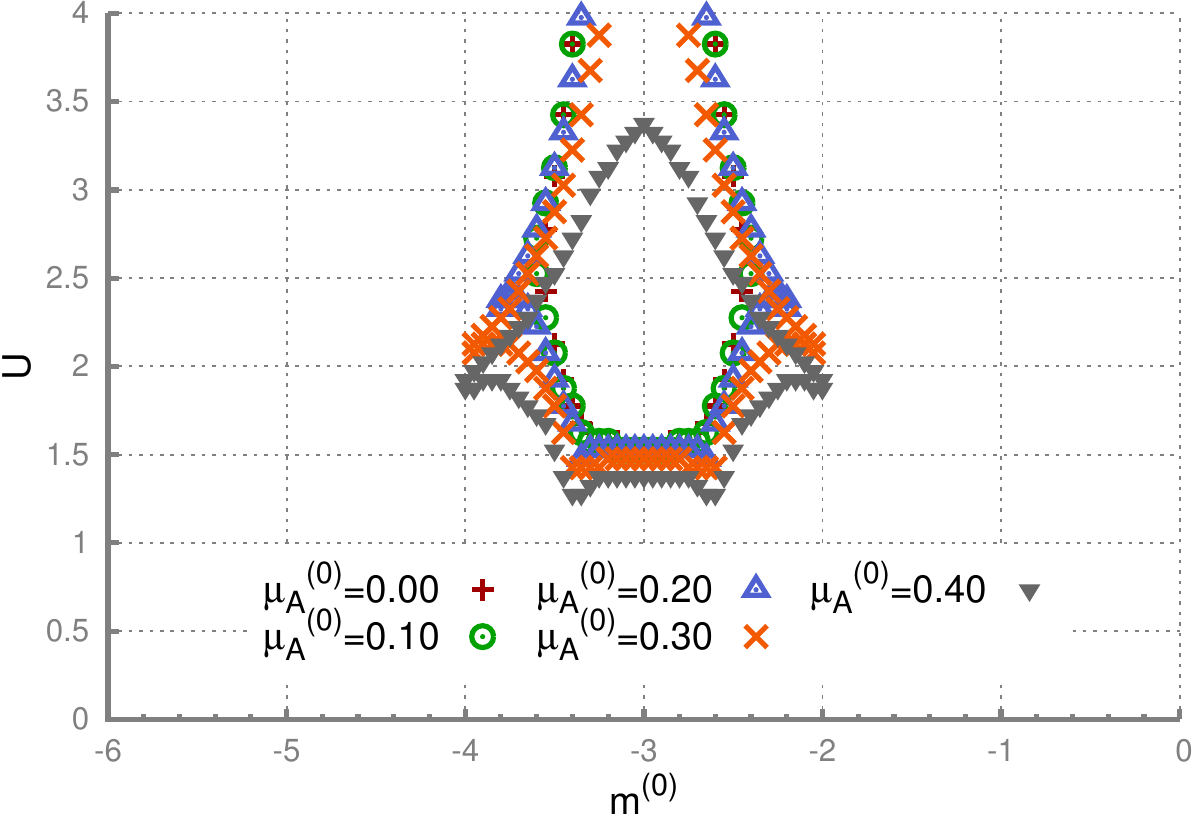}
  \caption{Mean-field phase diagram of our model in the $\mb$-$U$-plane for different values of the bare chiral chemical potential $\mua$. The points mark the boundaries of the Aoki phase. }
  \label{fig:phdiag_L50}
\end{figure}

 We plot the mean-field phase diagram of our model in the $\mb$-$U$-plane for different values of the bare chiral chemical potential $\mua$ on Fig.~\ref{fig:phdiag_L50}. The phase diagram is symmetric around the line of $\mb = -3.00$. Increasing the bare chiral chemical potential $\mub$ shifts the onset of the Aoki phase to slightly smaller values of the inter-electron interaction $U$. A similar effect of the chiral chemical potential on the critical interaction potential for spontaneous breaking of chiral symmetry has been found in \cite{Buividovich:14:1} for continuum Dirac fermions.

 In a previous study \cite{Buividovich:14:2}  we presented results for ${\ls = 10}$, since numerical calculations for $\ls = 8, 10, 12$ indicated that finite size effects have little influence on the mean-field phase diagram. This holds true in general but  we observe that the thin structures which point to the characteristic values of $\mb = 0,-2,-4,-6$ and are often called ``Aoki fingers'' get thinner as $\ls$ is increased. In the case of $\ls = 50$ we were no longer able to resolve the fingers numerically, see Fig.~\ref{fig:phdiag}.  Even for small lattice sizes the fingers are only present at zero bare chiral chemical potential and vanish as soon as $\mub \neq 0$. For the calculation of the chiral magnetic conductivity we are interested in values $0 < \mub < 0.3$, where the phase diagrams for $\ls=10$ and $\ls=50$ lie on top of each other. Therefore the volume dependence of the Aoki fingers at $\mub=0$ is of little interest for the main points of this paper and we discuss the origin of finite size effects in Appendix \ref{apdx:aoki}.

\begin{figure}[h!tpb]
  \centering
  \includegraphics[width=0.9\linewidth]{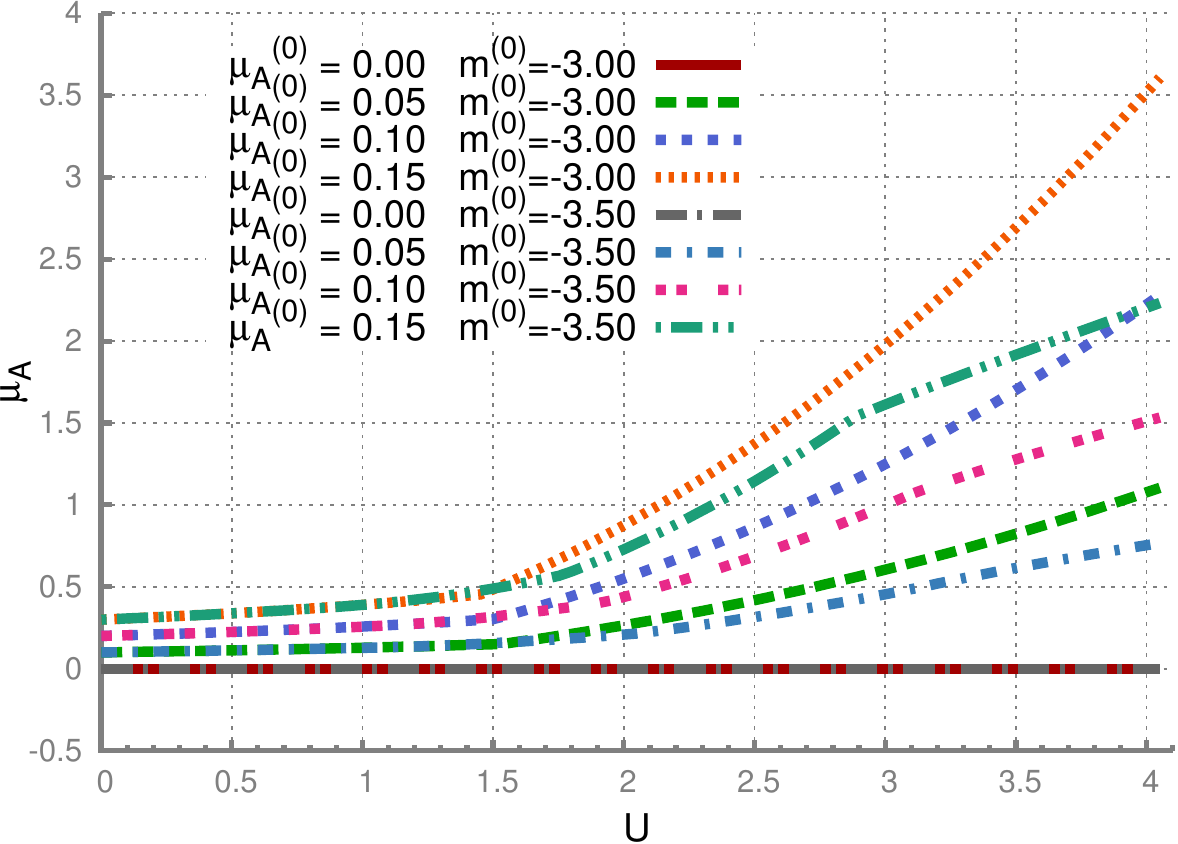}
  \caption{Renormalized chiral chemical potential $\mu_A$ as a function of the interaction potential $U$ for different values of the bare chiral chemical potential $\mub$.}
  \label{fig:ren_chem_pot}
\end{figure}

 Similarly to the case of continuum Dirac fermions \cite{Buividovich:14:1}, our lattice model also exhibits a strong renormalization of the chiral chemical potential. Figure \ref{fig:ren_chem_pot} shows the chiral chemical potential as a function of the inter-electron interaction $U$ for different values of the bare chiral chemical potential $\mub$. We note that the growth of $\mu_A$ with $U$ becomes somewhat more pronounced within the phase with spontaneously broken $\mathcal{C}\mathcal{P}$ symmetry.

 Let us now discuss in more detail the validity of our assumption of the spatial homogeneity of the fermionic condensates. In the process of calculating the current-current correlator (\ref{mf_response}) we have to compute the Hessian matrix $\frac{\partial^2 \mathcal{F}\lrs{\Phi_x, A_{x,k}}}{\partial \Phi_{x, A} \partial \Phi_{y, B}}$ of second derivatives of the mean-field free energy with respect to the Hubbard-Stratonovich field $\Phi_{x, A}$ in the vicinity of the homogeneous configuration $\Phi_x = \Phi^{\star}$ which minimizes the mean-field free energy functional (\ref{free_energy_mf1}). If the homogeneous configuration $\Phi_x = \Phi^{\star}$ is a local minimum in the space of Hubbard-Stratonovich fields $\Phi_{x,A}$, all the eigenvalues of this matrix should be positive. Since $\Phi^{\star}$ is a spatially homogeneous field configuration, the entries of this Hessian matrix depend only on $x - y$, and we can calculate its eigenvalues by performing the Fourier transformation with respect to $x - y$. At each wave vector in the Fourier transform, one then has to diagonalize only a $15 \times 15$ matrix with entries corresponding to different spinor structures of $\Phi_x$ (see (\ref{Phi_basis_decomposition})). We have calculated the eigenvalues of the Hessian matrix of $\mathcal{F}\lrs{\Phi}$ for every point in the parameter space which we have considered in our mean-field calculations and for wave vectors $k_3 = 2\pi m/\ls, \ m \in \{0,\dots,8\}$ and found that they are always positive. This calculation suggests that the uniform saddle-point configuration $\Phi^{\star}$ of the Hubbard-Stratonovich field defined in (\ref{restr_saddle_point}) is indeed a local minimum in the space of all possible field configurations of $\Phi_x$ and if there is a non-homogeneous field configuration which further decreases the free energy it should be separated from the homogeneous configuration $\Phi^{\star}$ by some potential barrier. These findings to some extend justify the assumption of homogeneity made for the saddle point value of the field. In the concluding Section \ref{sec:conclusions}, we further discuss the plausibility of this assumption in view of some recent theoretical studies which predict inhomogeneous ground state for chirally imbalanced fermions.

\begin{figure}[h!tpb]
  \centering
  \includegraphics[width=0.9\linewidth]{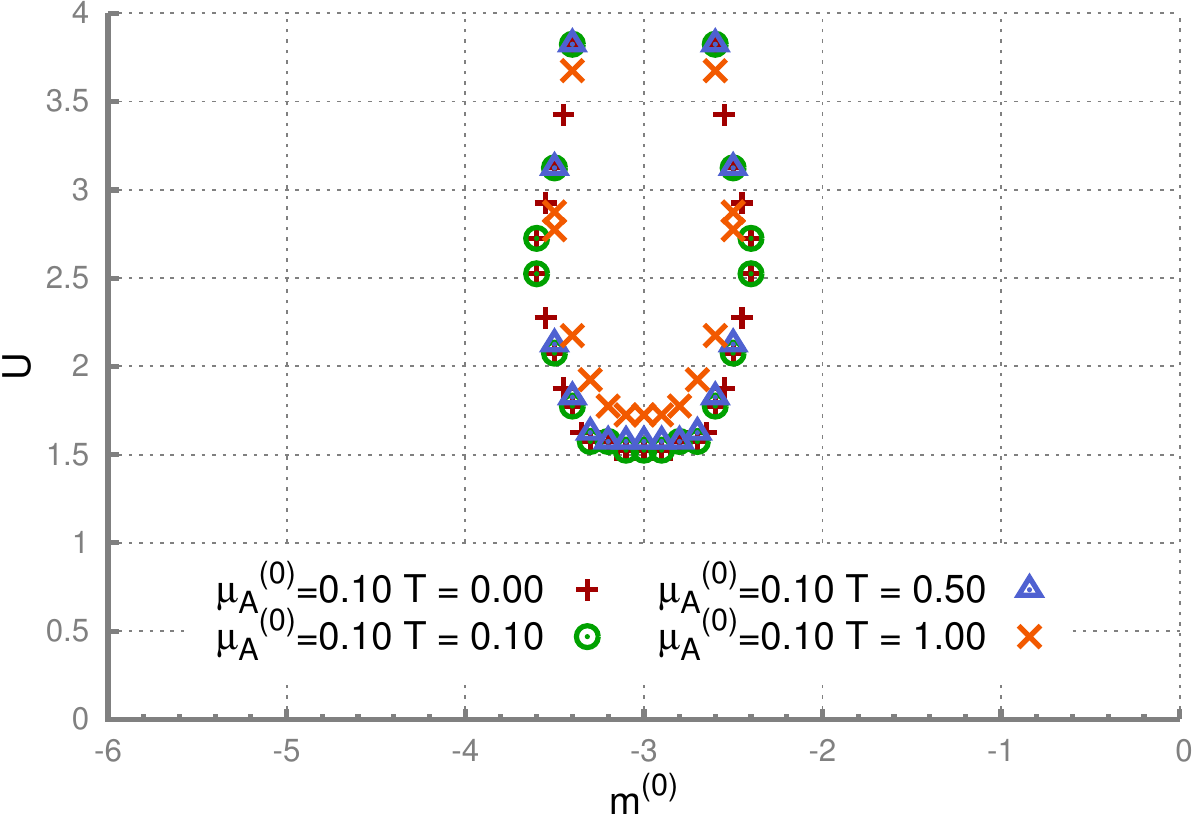}\\
  \caption{Phase diagram in the parameter space of bare mass $\mb$ and the interaction potential $U$ at different temperatures. The chiral chemical potential is fixed at $\mua^{\lr{0}} = 0.1$.}
  \label{fig:phdiag_finite_T}
\end{figure}

 Finally, we briefly consider the effect of finite temperature on the phase structure of our model. On Fig.~\ref{fig:phdiag_finite_T} we plot the boundaries of the Aoki phase in the parameter space of bare mass $\mb$ and the interaction potential $U$ at different temperatures. The bare chiral chemical potential is fixed at $\mua^{\lr{0}} = 0.1$. As expected, finite temperature tends to destroy the fermionic condensates and hence shifts the phase with spontaneously broken $\mathcal{C}\mathcal{P}$ symmetry towards stronger interactions. This shift is, however, quite small even for temperature $T = 1$ comparable to the band width in our model. We conclude therefore that the effect of temperature is rather small at the level of the phase diagram.

\section{Mean-field calculation of the static chiral magnetic conductivity}
\label{sec:cmc}

\begin{figure*}[h!tpb]
  \centering
  \includegraphics[width=0.45\linewidth]{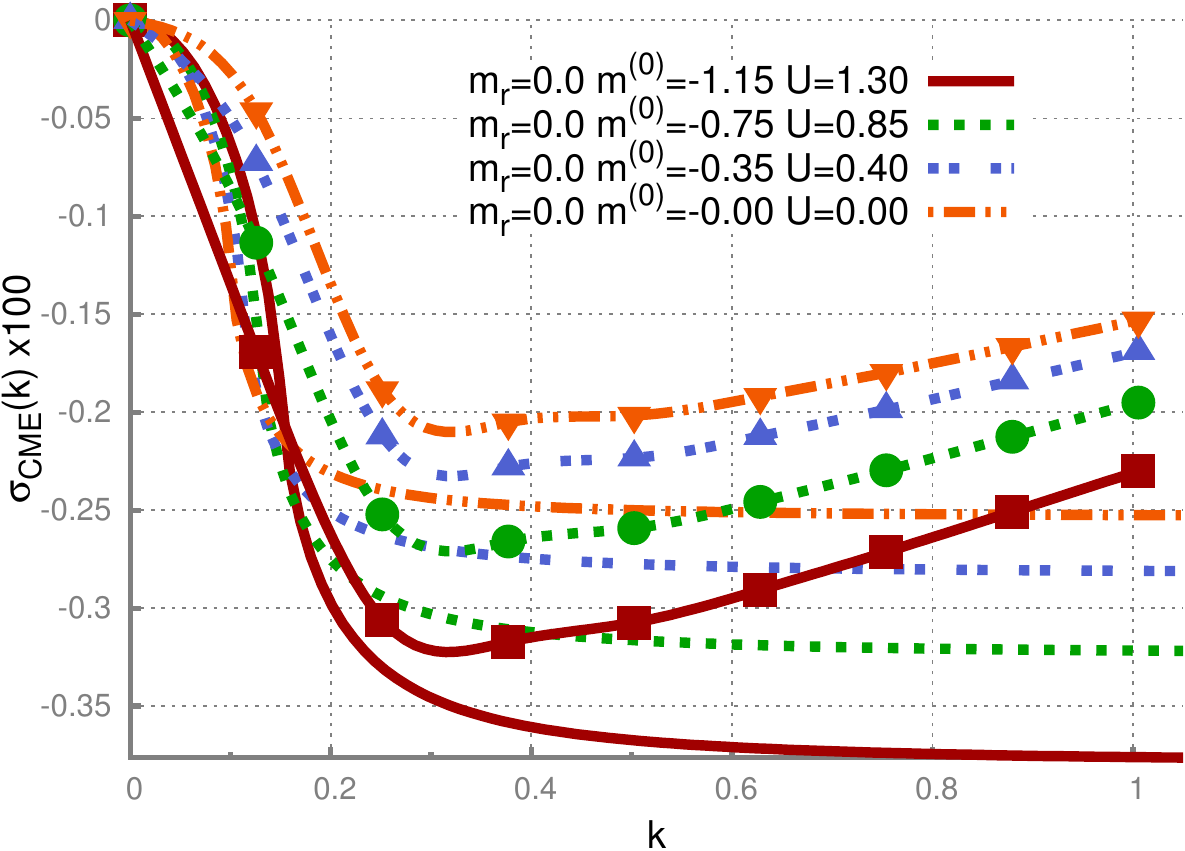}
  \includegraphics[width=0.45\linewidth]{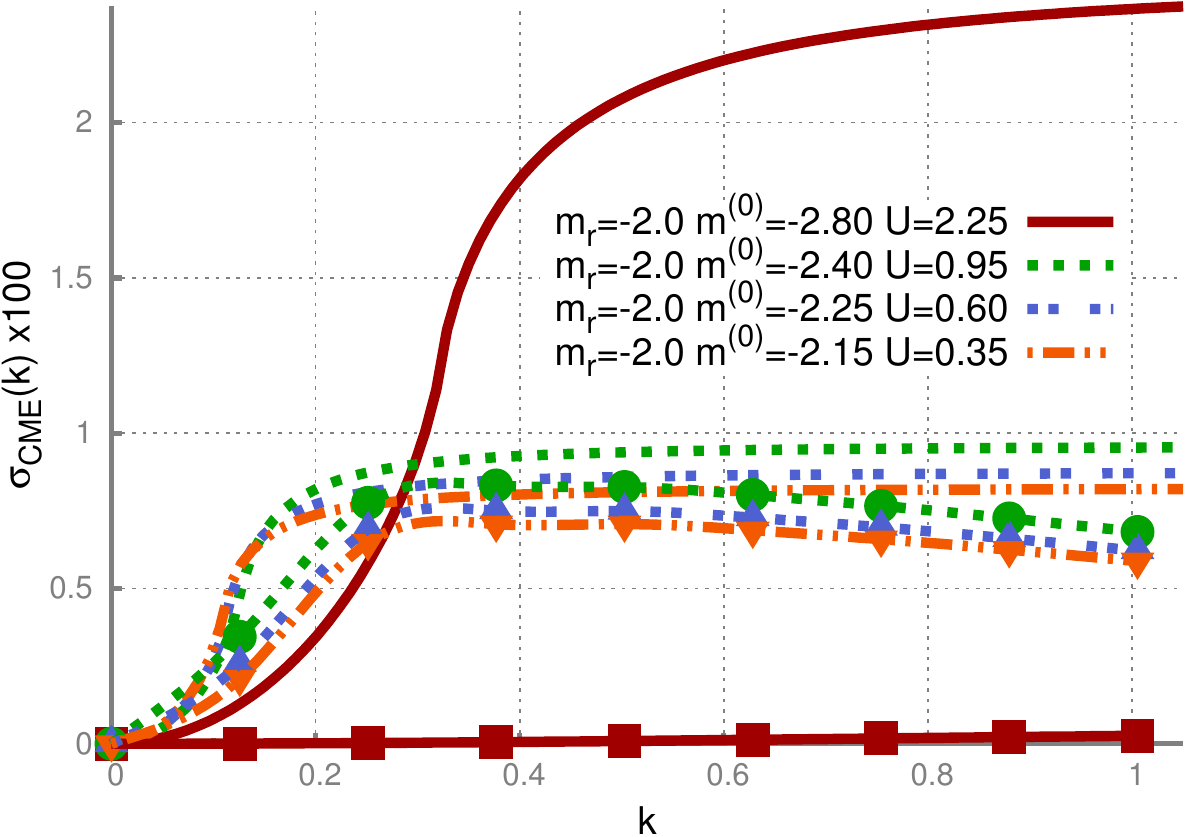}\\
  \includegraphics[width=0.45\linewidth]{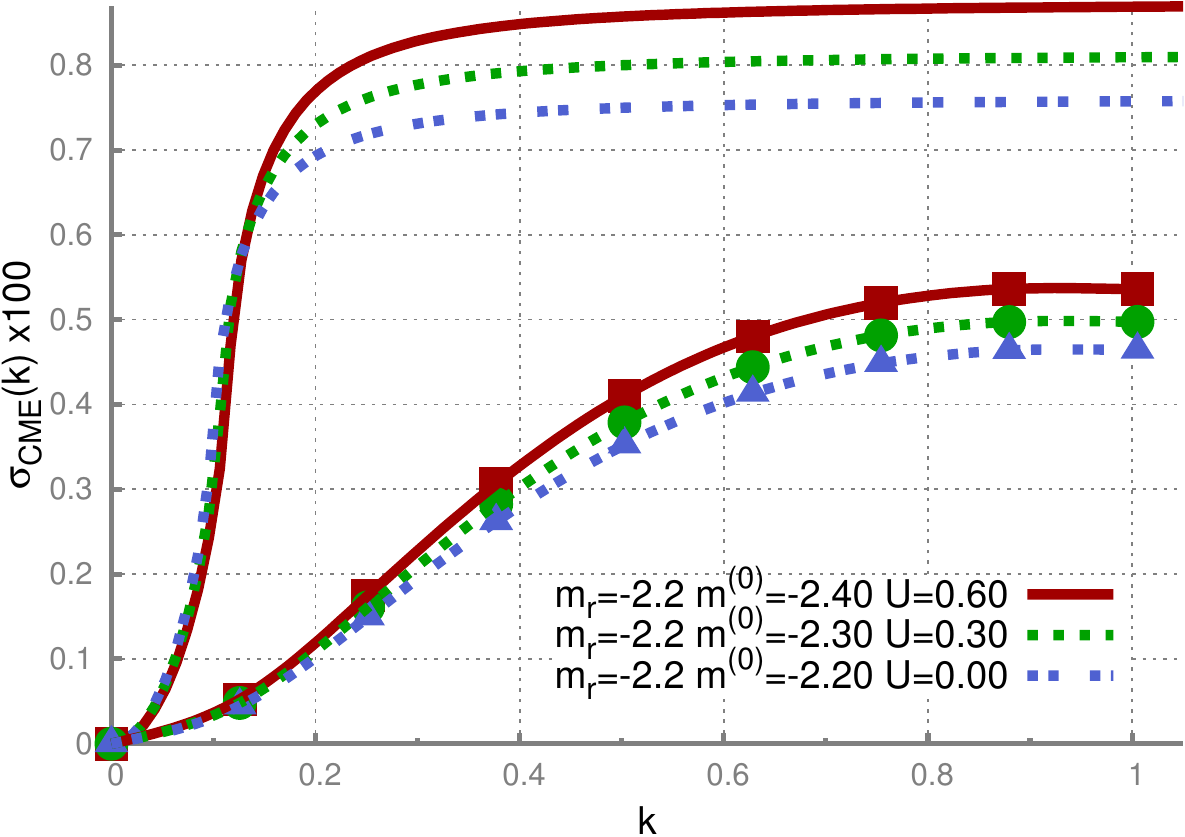}
  \includegraphics[width=0.45\linewidth]{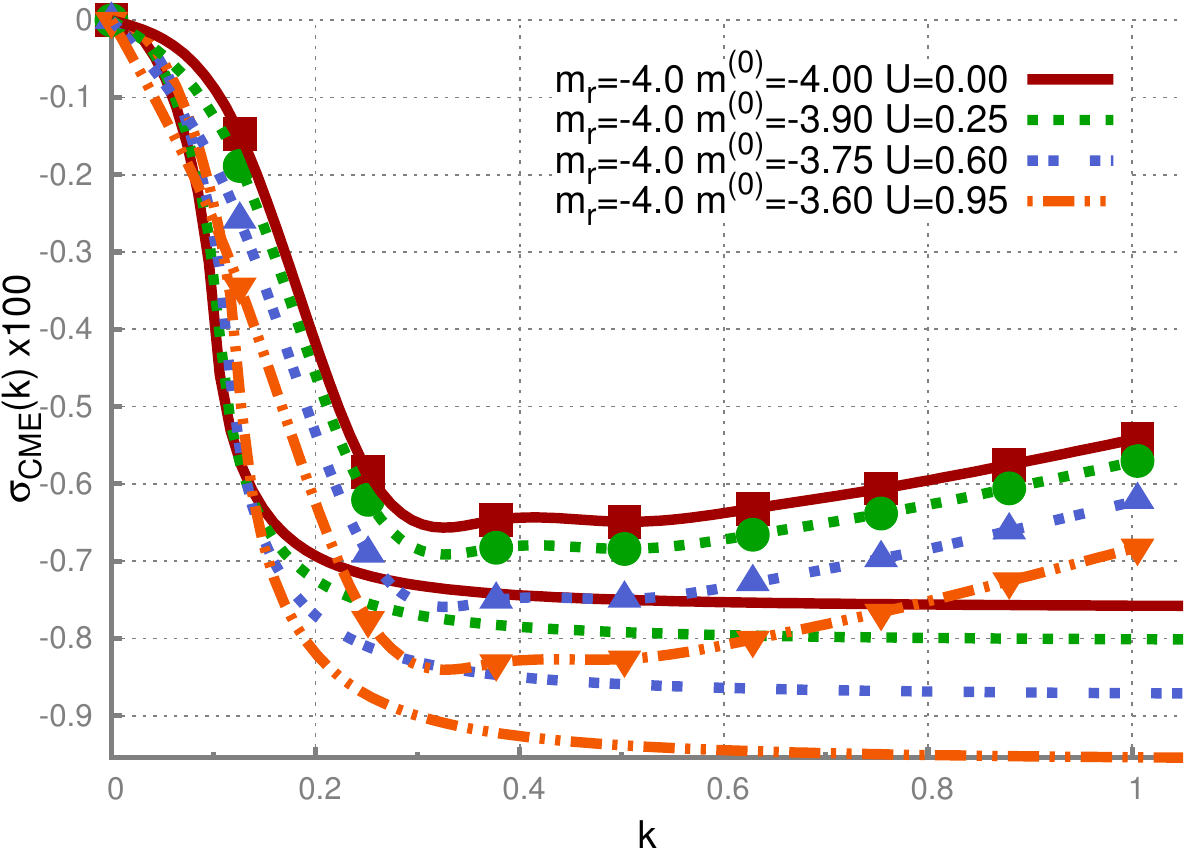}
  \caption{Numerical results for the chiral magnetic conductivity $\sigma_{CME}\lr{k}$ at characteristic values of $\mr$ for different values of the interaction potential $U$ (lines with points). Clockwise from top left: $\mr = 0.00, \mr = -2.00, \mr = -4.00$ and $\mr = -2.20$. In all cases $\mub = 0.05$ and $L_s = 50$. For comparison we also plot the result (\ref{eq:ft_current_current_PV}) for free continuum Dirac fermions with Pauli-Villars regularization and with the renormalized chiral chemical potential $\mua$ (lines of the same style as for numerical data, but with no points on them). }
  \label{fig:JJ_big}
\end{figure*}

 Once we have found the renormalized values of $m_r$, $m_i$ and $\mua$ for our model, they can be used to calculate the chiral magnetic conductivity $\scme\lr{k}$ from equations (\ref{mf_response}) and (\ref{cmc_Kubo}).  On Fig.~\ref{fig:JJ_big} we plot $\scme\lr{k}$ as a function of the wave vector $\vec{k}$ (which is parallel to one of the coordinate axes) of the external magnetic field at several characteristic points in the phase diagram of our model. The bare chiral chemical potential is $\mu_A^{\lr{0}} = 0.05$ everywhere. Within each plot, we scan the phase diagram along the line of constant renormalized mass $m_r$. Since for Dirac/Weyl semimetals one is typically interested in the regions with gapless spectrum, we have considered the values $m_r = 0$, $m_r=-2$ and $m_r = -4$. According to the discussion in Section \ref{sec:phase_diagram} above, at $m_r = -2, -4$ and $m_r = 0, -6$ the dispersion relation of our model has $N_f = 3$ or $N_f = 1$ Dirac cones, correspondingly. This results in an obvious linear scaling of $\scme\lr{k}$ with $N_f$, which can be clearly seen on Fig.~\ref{fig:JJ_big}. Moreover, one can show that the renormalized masses $m_r$ and $-6 - m_r$ are equivalent, but the roles of left- and right-handed spinor components interchange between them. In other words, the chiral chemical potential effectively changes sign between $m_r$ and $-6 - m_r$. This can be clearly seen on Fig.~\ref{fig:JJ_big} for the lines with $m_r = -2$ and $m_r = -4$, for which $\scme\lr{k}$ only differs by a sign. We note also that by virtue of such change of sign between the values $m_r$ and $-6 - m_r$ the chiral magnetic conductivity $\scme\lr{k}$ exactly vanishes for all $k$ on the line $m_r = -3$, which is the symmetry axis of the phase diagram of our model.

 With these model-dependent features taken into account, we observe a qualitative agreement between our data at small $k$ outside of the Aoki phase and the continuum result for $\scme\lr{k}$ obtained with Pauli-Villars regularization \cite{Buividovich:13:8,Ren:11:1}:
\begin{eqnarray}
\label{eq:ft_current_current_PV}
 \scme^{PV}\lr{k}
 = \nonumber \\ =
 \frac{N_f}{\lr{2\pi}^2}\lr{\mua + \frac{\mua^2 - k^2/4}{k} \log\left|\frac{2\mua - k}{2\mua + k} \right|} .
\end{eqnarray}

 In particular, one can see the characteristic growth of $\scme\lr{k}$ around $k = 2 \mua$. At $k > 2 \mua$ $\scme\lr{k}$ approaches the asymptotic value $\pm \frac{\mua}{2 \pi^2}$ of the continuum result, but at even larger $k$ we observe a decrease due to explicit chiral symmetry breaking at larger momenta. It is important to stress that the behavior of $\scme\lr{k}$ depends on the renormalized chiral chemical potential $\mua$, which is an experimentally measurable quantity. E.g. the angular-resolved photoemission spectroscopy (ARPES) measurements in the vicinity of the Dirac cones (see for example \cite{Kharzeev:14:1}) should yield the renormalized chiral chemical potential $\mua$. Therefore we use the renormalized value of $\mu_A$ in order to compare our data with the continuum result (\ref{eq:ft_current_current_PV}).

 We see that for all plots the lattice result for $\scme\lr{k}$ is smaller than the continuum result (\ref{eq:ft_current_current_PV}) calculated with the renormalized chiral chemical potential. We conclude therefore that the observed enhancement of $\scme\lr{k}$ is mostly due to the growth of $\mu_A$ with interaction strength. As the effective single-particle Hamiltonian becomes gapped due to a nonzero $\mathcal{C}\mathcal{P}$-breaking mass term in the Aoki phase, $\scme\lr{k}$ becomes significantly suppressed at all $k$, despite the even stronger growth of $\mu_A$ in this phase. In order to illustrate the behaviour of $\scme\lr{k}$ in the case when the spectrum of the effective single-particle Hamiltonian is gapped outside of Aoki phase, on Fig. \ref{fig:JJ_big} (at the bottom on the left) we also plot it along the line with $m_r = -2.20$. We again see that the CME response becomes suppressed, as could be expected also from continuum calculations \cite{Buividovich:13:8,Ren:11:1}.

 Let us now check whether the behavior which we have found for the above specific values of model parameters is typical throughout the whole phase diagram of our model. To this end, it is advantageous to characterize $\scme\lr{k}$ in terms of its asymptotic value at $\mu_A \ll k \ll 1$. In this range of momenta the lattice result for $\sigma_{CME}\lr{k}$ should approach the continuum one, which tends to the asymptotic value $\frac{N_f \mua}{2 \pi^2}$ at $k \gg \mu_A$ (see Fig.~\ref{fig:JJ_big}). The universality of this asymptotic value is due to its relation to the anomaly coefficient, which is, however, quite nontrivial \cite{Buividovich:13:8}.

 However, our calculations are performed on a finite lattice, and the values of $k$ are quantized. Therefore we have a rather narrow range of lattice momenta $\mu_A \ll k \ll 1$ for which it makes sense to compare the chiral magnetic conductivity $\scme\lr{k}$ with $\frac{N_f \, \mu_A}{2 \pi^2}$. We have chosen to estimate the asymptotic value $\sigma\lr{k}$ at $\mua \ll k \ll 1$ simply as the maximal value of $\sigma_{CME}\lr{k}$ among all discrete momenta which we consider. We denote this estimate as $\smax$:
\begin{eqnarray}
\label{smax_def}
 \smax = \max\limits_k \, \sigma_{CME}\lr{k} .
\end{eqnarray}

\begin{figure}[h!tpb]
  \centering
  \includegraphics[width=0.9\linewidth]{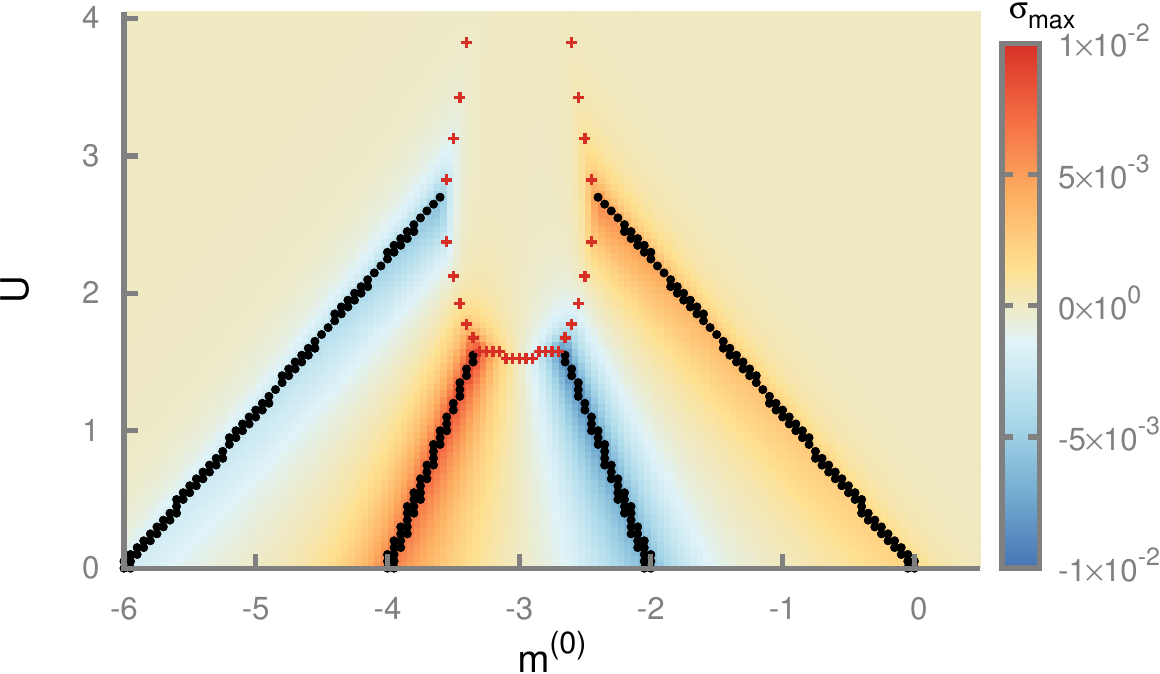}
  \caption{Numerical estimate $\smax$ of the asymptotic value of $\scme\lr{k}$ at $\mu_A \ll k \ll 1$ as a function of bare mass $\mb$ and the interaction potential $U$ at fixed bare chiral chemical potential $\mu_A^{\lr{0}} = 0.05$. The red crosses mark the border of the Aoki phase and the black points mark the lines of constant $m_r = 0, \, -2, \, -4, \, -6$.}
  \label{fig:s0_mu005}
\end{figure}

 On Fig.~\ref{fig:s0_mu005} we plot our numerical estimate $\smax$ of $\scme\lr{k}$ at $\mu_A \ll k \ll 1$ in the parameter space of the bare mass $\mb$ and the interaction potential $U$ for $\mu_A^{\lr{0}} = 0.05$. In agreement with our previous observations, we find that $\sigma_{max}$ is peaked around the lines of constant renormalized mass $m_r = 0, \, -2, \, -4, \, -6$, at which our model has massless excitations. In the gapped phase with spontaneously broken $\mathcal{C}\mathcal{P}$ symmetry, $\smax$ quickly decreases. In general, the behavior of $\smax$ is non-trivial and depending on the location on the bare mass axis increasing the inter-electron interaction strength $U$ can lead to an increase or a decrease of $\smax$.

\begin{figure}[h!tpb]
  \centering
  \includegraphics[width=0.9\linewidth]{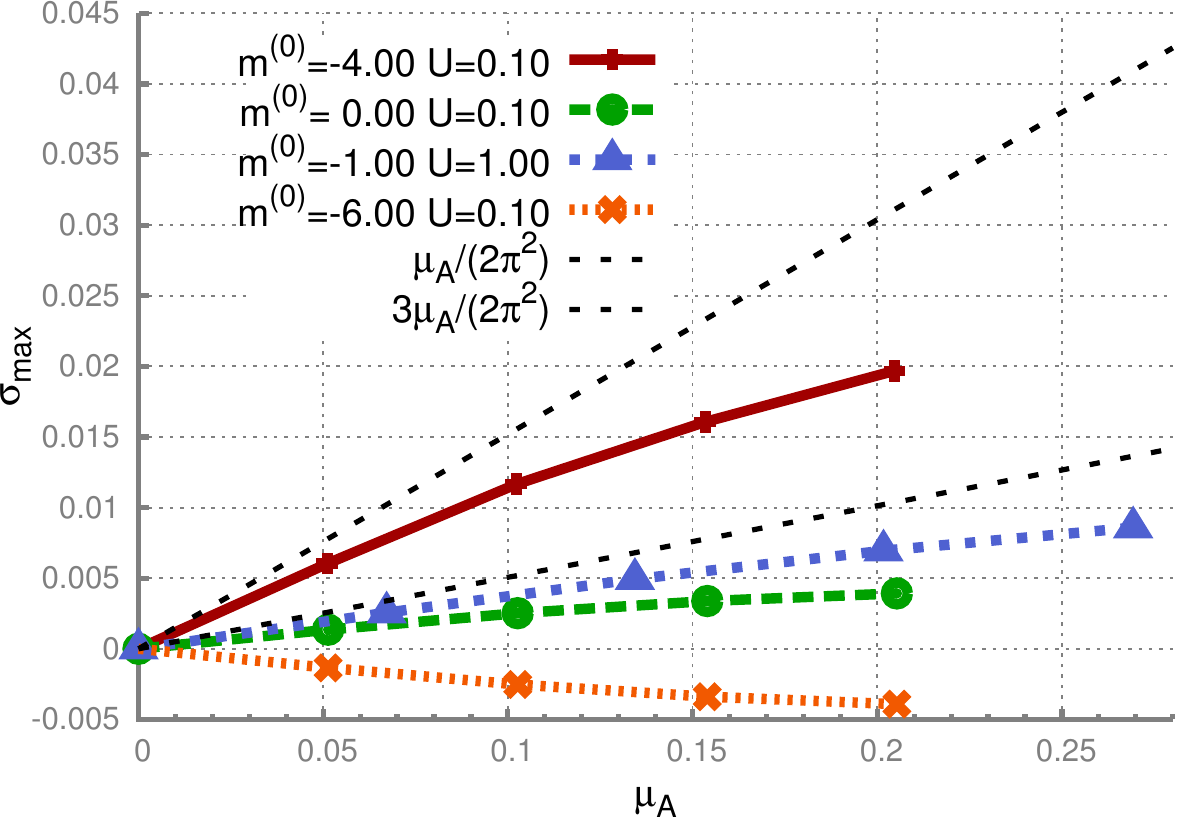}
\caption{Numerical estimate $\smax$ of the asymptotic values of $\scme\lr{k}$ at $\mu_A \ll k \ll 1$ as a function of the renormalized chiral chemical potential $\mua$ for different values of the bare mass and the interaction potential. The points are connected with straight lines to guide the eye. For small $\mua$ $\smax$ is a linear function of the renormalized chiral chemical potential.  }
\label{fig:lin_slope}
\end{figure}

 In order to compare our estimate of the asymptotic values of $\scme\lr{k}$ at $\mua \ll k \ll 1$ with the universal value $\frac{\mu_A}{2 \pi^2}$ in (\ref{cmc_theory0}) which is related to axial anomaly coefficient, we further assume that $\smax$ is a linear function of $\mua$. In order to justify this assumption, on Fig.~\ref{fig:lin_slope} we show the dependence of $\smax$ on $\mua$ for several different points in the parameter space of our model. We thus estimate the ratio of the asymptotic value of $\scme\lr{k}$ at $\mu_A \ll k \ll 1$ to the renormalized chiral chemical potential $\mua$ by fitting $\smax$ at $\mub=0.05, \, 0.10, \, 0.15$ with a linear function $\smax = A \, \mu_A$, where $A$ is the fitting parameter. We note also that in realistic experiments, it is difficult to achieve large values of $\mua$ comparable to the band width, so in practice one is anyway interested in small values of $\mua$.

\begin{figure}[h!tpb]
  \centering
  \includegraphics[width=0.9\linewidth]{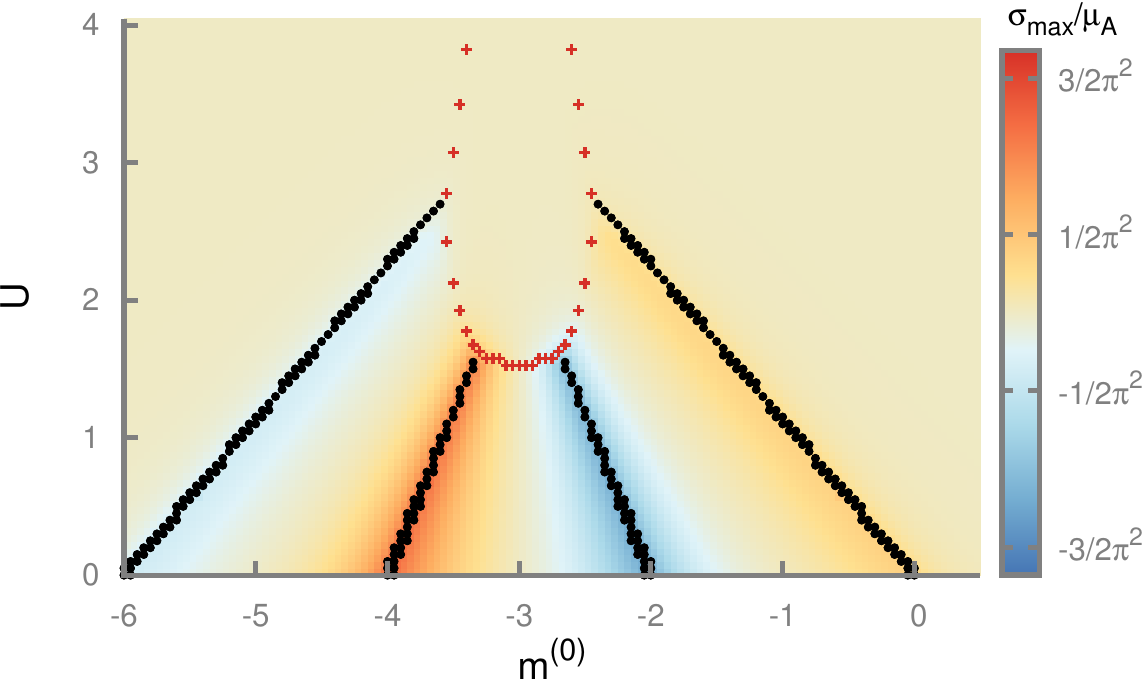}
  \caption{The ratio of the estimated asymptotic value $\smax$ of $\scme\lr{k}$ at $\mua \ll k \ll 1$ to the renormalized chiral chemical potential $\mua$. The black dots mark the lines of constant renormalized mass $\mr = -6, \, -4, \, -2$ and $m_r = 0$, respectively, and the red crosses indicate the border of the Aoki phase for $\mub=0.0$.}
  \label{fig:s0_fit_mr}
\end{figure}

 Our numerical estimate $\frac{\smax}{\mua}$ of the ratio $\frac{\scme\lr{k}}{\mua}$ at $\mua \ll k \ll 1$ is shown on Fig.~\ref{fig:s0_fit_mr} as a function of the bare mass $\mb$ and the interaction potential $U$. Just as on Fig.~\ref{fig:s0_mu005} we see that this ratio is peaked along the lines of constant $\mr = 0, -2, -4$ and $-6$, where the fermionic single-particle spectrum is gapless, and is strongly suppressed in the vicinity of the Aoki phase. We see again that even in the vicinity of the lines with $N_f = 1$ and $N_f = 3$ massless Dirac cones in the energy spectrum the ratio $\frac{\smax}{\mua}$ never exceeds the universal value $N_f/\lr{2 \pi^2}$. We thus conclude that strong enhancement of $\scme$ observed for the continuum Dirac fermions with spontaneously broken chiral symmetry in \cite{Buividovich:14:1} does not happen in the lattice model with only emergent chiral symmetry.

\begin{figure*}[h!tpb]
  \centering
  \includegraphics[width=0.45\linewidth]{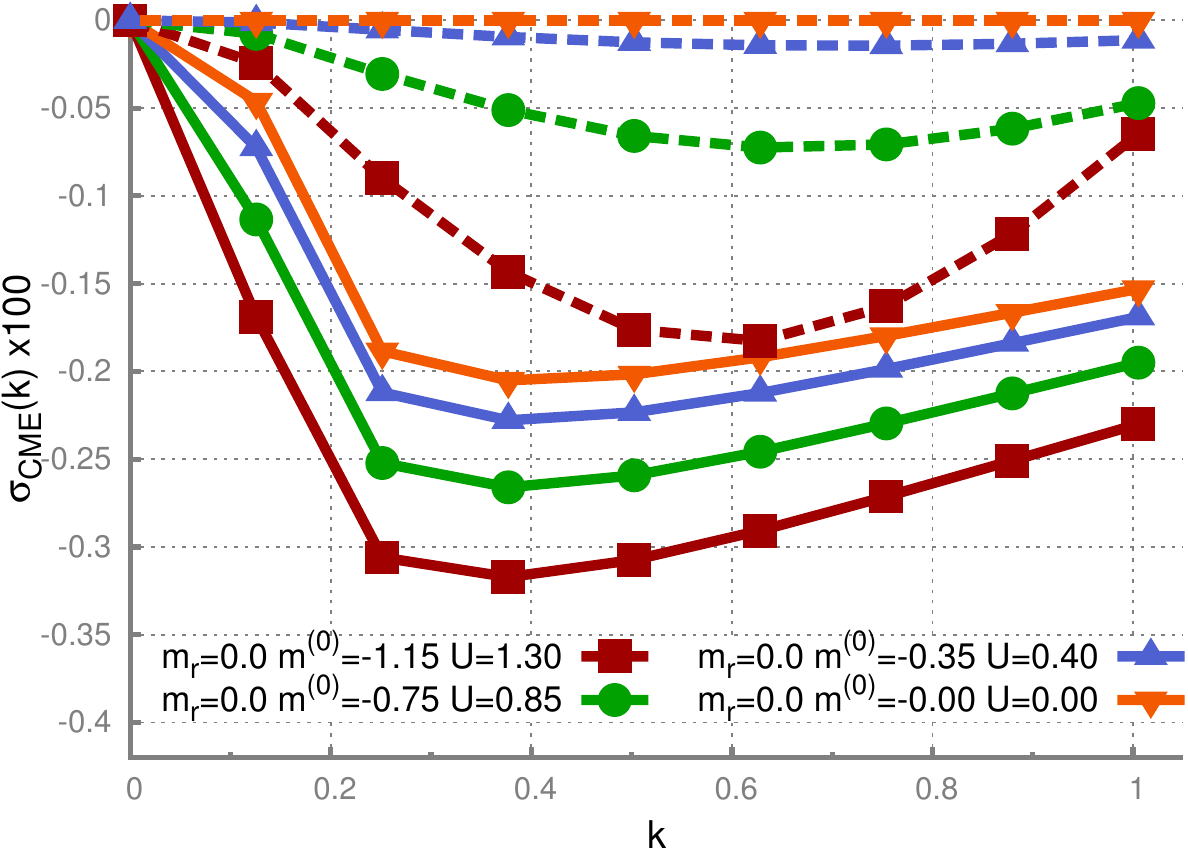}
  \includegraphics[width=0.45\linewidth]{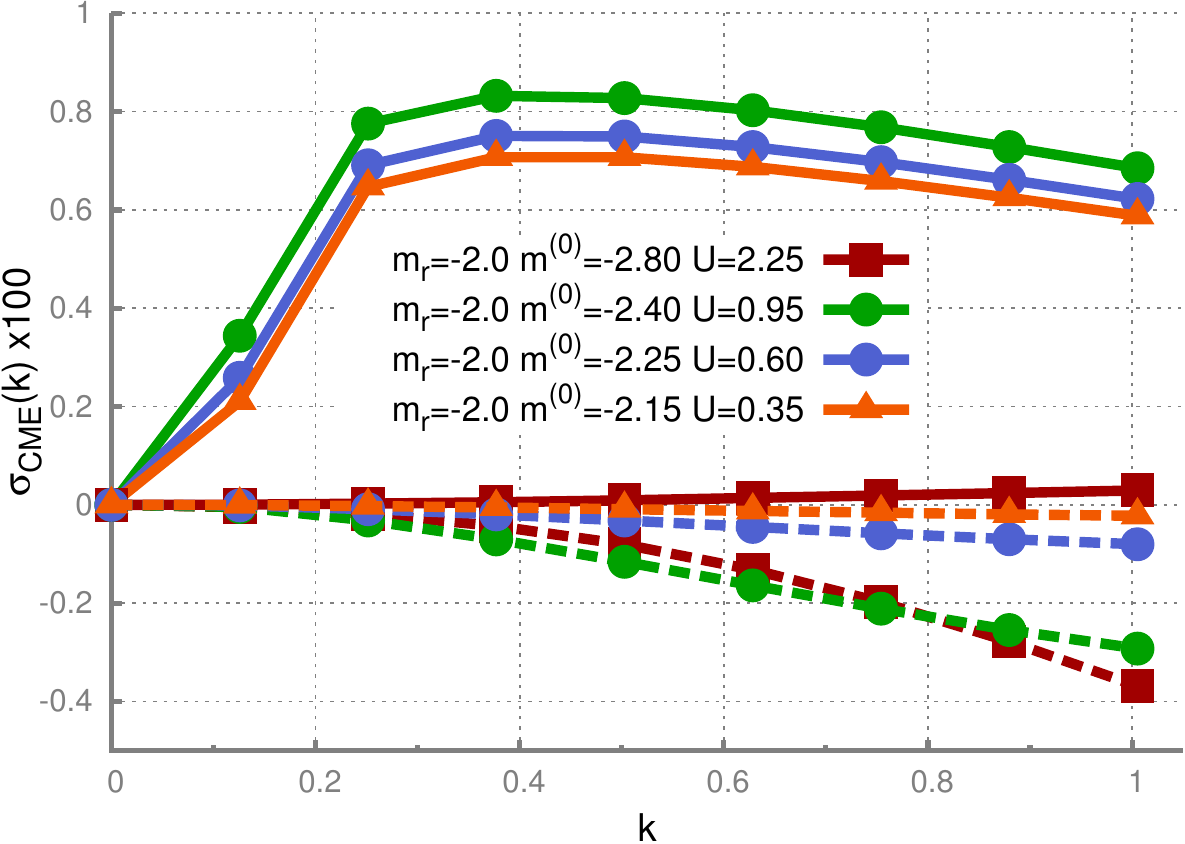}\\
  \caption{Relative importance of nontrivial corrections to $\scme\lr{k}$ due to inter-fermion interactions. Solid lines with symbols correspond to the contribution of tree diagram to $\scme\lr{k}$ (first summand in (\ref{mf_response})) and dashed lines with the same symbols - to the contribution of loop diagrams (second summand in (\ref{mf_response})). Since the contribution of loop diagrams turns out to be numerically very small, we multiply it by a factor $10^2$ ($\mr = -2.0$) or $10^4$ ($\mr = 0.0$).}
  \label{fig:loop_contr}
\end{figure*}

 It is also interesting to check what is the nontrivial contribution of inter-electron interactions to $\scme\lr{k}$. As discussed in Section \ref{sec:mean_field_general}, in the general expression (\ref{mf_response}) for the current-current correlators the first, tree-level, summand describes the change of the electromagnetic response merely due to renormalization of the parameters $\mua$, $m_r$ and $m_i$ of the effective single-particle Hamiltonian, whereas the second summand describes nontrivial loop corrections originating from ladder diagrams with arbitrary number of fermionic loops. In order to quantify the importance of interactions, on Fig.~\ref{fig:loop_contr} we separately plot the tree diagram contribution to $\scme\lr{k}$ (first summand in (\ref{mf_response})) and the contribution of loop diagrams (second summand in (\ref{mf_response})) as a function of $k_3$ for $m_r = 0$, $\mub = 0.05$ and $m_r = -2$, $\mub = 0.05$. We see that the loop contributions are numerically very small in the phase with unbroken $\mathcal{C}\mathcal{P}$ symmetry - at maximum around one percent for $m_r = -2$, and by almost two orders of magnitude less for $m_r = 0$. It is interesting to note here the dependence on the number of Dirac cones in the spectrum. The only exception is the point with $m_r = -2$ and $U = 2.25$ in the phase with broken $\mathcal{C}\mathcal{P}$ symmetry, where the contribution of the tree diagram becomes very small due to gap opening, but the loop contributions stay almost the same as in the unbroken phase, thus they become more important.

\begin{figure}[h!tpb]
  \centering
  \includegraphics[width=0.9\linewidth]{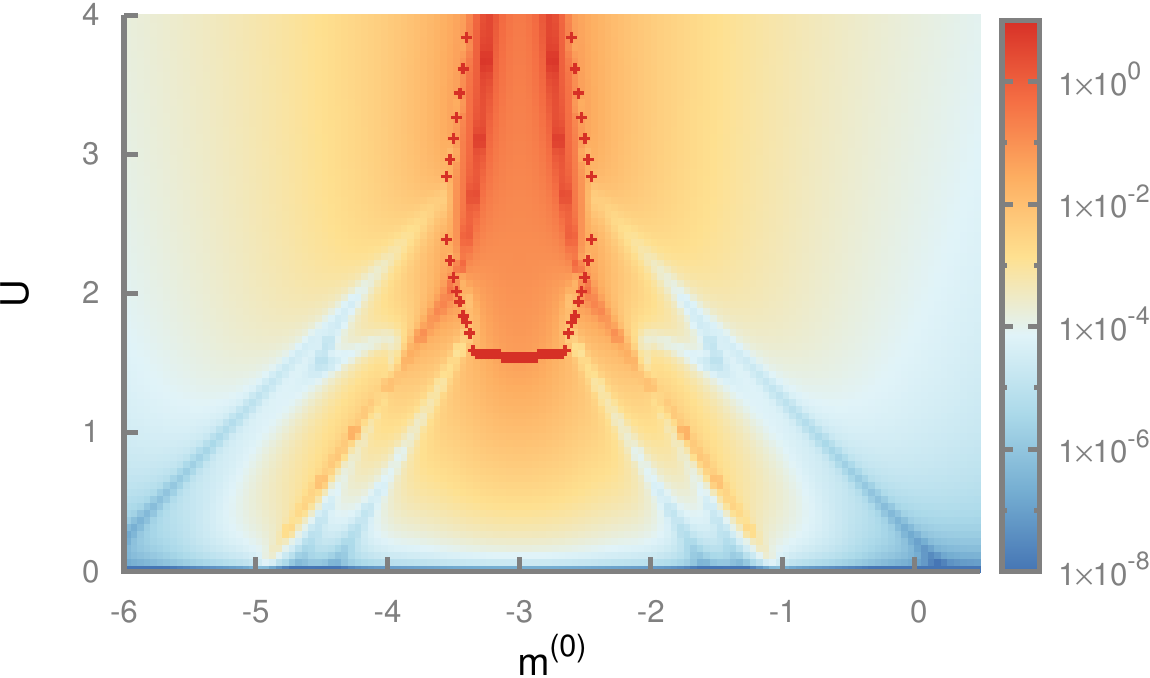}
  \caption{The ratio of the contribution of tree-level diagram (first term in (\ref{mf_response})) to the contribution of the loop diagrams (second term in (\ref{mf_response})) to the chiral magnetic conductivity $\scme\lr{k}$ at $k = 6$ and $\mub = 0.05$ as a function of the bare mass $\mb$ and interaction potential $U$.}
  \label{fig:JJ_comp_005}
\end{figure}

 We further illustrate the importance of loop contributions on Fig.~\ref{fig:JJ_comp_005} by plotting the ratio of the tree diagram and the loop contributions to $\scme\lr{k}$ at fixed $k = 6$ as a function of the bare mass $\mb$ and the interaction potential $U$. We see again that the loop contribution is very small along the lines with massless Dirac cones in the spectrum, and only becomes important in the regions of phase diagram where the energy spectrum has a large gap and thus the tree diagram contribution is strongly suppressed. Within the ``axionic insulator'' phase with broken $\mathcal{C}\mathcal{P}$ symmetry, the importance of loop contributions reaches its maximum. In particular, it is interesting to note that the loop contribution is strongly peaked along the two lines starting roughly from $\mb = -5$ and $\mb = -1$ and going into the region of the broken $\mathcal{C}\mathcal{P}$ phase.

\begin{figure*}[h!tpb]
  \centering
  \includegraphics[width=0.45\linewidth]{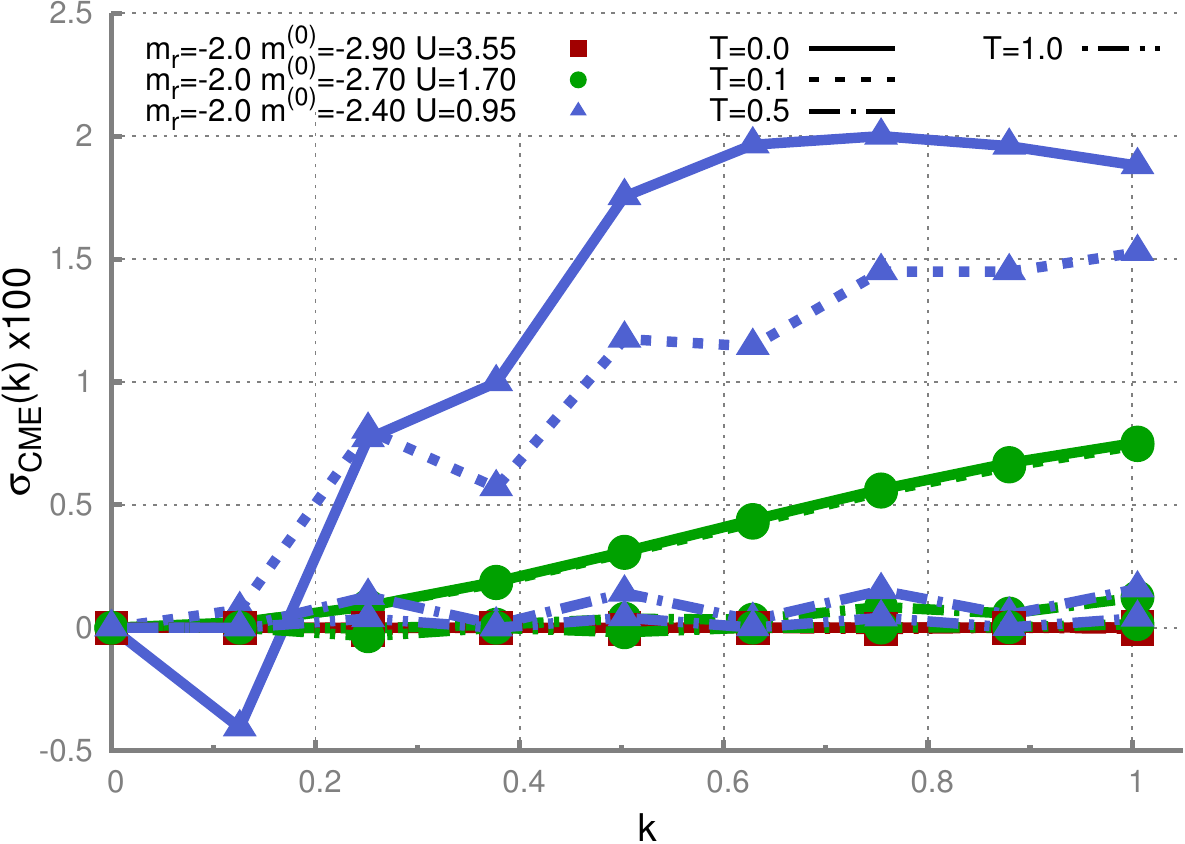}
  \includegraphics[width=0.45\linewidth]{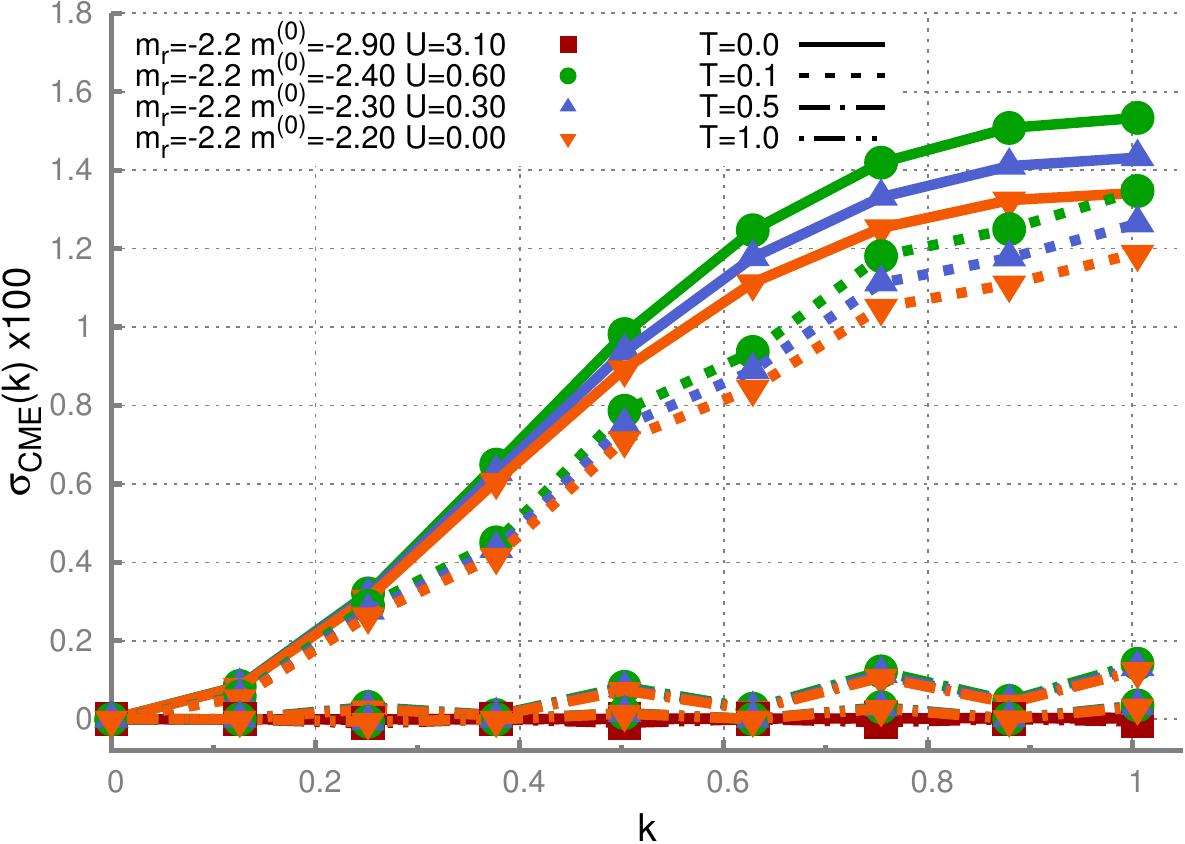}\\
  \caption{Numerical results for $\scme\lr{k}$ at finite temperatures. Point styles denote different sets of model parameters ($m_r$ and $U$) and line styles different values of temperature. The bare chiral chemical potential is $\mua^{\lr{0}} = 0.15$ everywhere.}
  \label{fig:cme_fintemp}
\end{figure*}

 Finally, on Fig.~\ref{fig:cme_fintemp} we illustrate the effect of finite temperature on the chiral magnetic conductivity $\scme\lr{k}$. For all values of the model parameters finite temperature suppresses the CME response. The suppression is not very large if the temperature is comparable with the chiral chemical potential. At higher temperatures of the order of the band width, $\scme\lr{k}$ becomes almost zero. Such temperature dependence of the chiral magnetic conductivity is expectable, since the temperature tends to reduce the differences in the occupation numbers of left- and right-handed fermions of the same momentum. Indeed, in the limit of infinite temperature, the Fermi distributions in (\ref{free_energy_fermion_gas}) and (\ref{ffe_2nd_der_general}) become independent of $\mua$, and hence the CME response should vanish.

\section{Connection with experiment}
\label{sec:experiment}

 Spatially modulated magnetic fields are very convenient for theoretical considerations of the chiral magnetic effect, in particular, for establishing its connection with the anomaly coefficient. In experiments, however, one typically subjects the Weyl semimetal sample to a static and constant magnetic field. Since our calculations show that $\scme(k)$ vanishes in the limit $k \to 0$, a natural question is how to measure such a quantity using a constant magnetic field. While the static Kubo formula (\ref{cmc_Kubo}) on which we base our calculations probably does not take into account all the details of the realistic experimental setup, in this Section we show that nevertheless our results provide a reasonable estimate of the CME response in experiments, at least by order of magnitude.

 To this end we note that the real samples of Weyl/Dirac semimetals used in experiments have finite size of order of $1 \textrm{mm}$ \cite{Xiong:15:1,Kharzeev:14:1}. Correspondingly, one can expect that magnetic fields have spatial modulation with characteristic wavelength of the same order or even shorter, due to inhomogeneities and grain structure of real crystals.

 In this paper, we have performed calculations for finite-size systems with periodic boundary conditions, which do not allow for arbitrarily small constant magnetic fields due to magnetic flux quantization constraints. On the other hand, outside of the axionic insulator phase and for sufficiently small momenta our results are quite well described by the expression (\ref{eq:ft_current_current_PV}) obtained for Dirac fermions with Pauli-Villars regularization in infinite continuum space. In order to model the spatial modulation of magnetic field due to the finite sample size in a way which is consistent with the approximations used in this paper, let us study the current in the case of a magnetic field that is static and constant inside a finite region and vanishes outside of this region. For simplicity we consider the magnetic field of an infinitely long solenoid with radius $R$ that is oriented in the $x_3$-direction. The magnetic field is constant inside the solenoid, points in $x_3$-direction and vanishes outside of the solenoid:
\begin{equation}
  \label{eq:B_solenoid}
   \vec{B}\lr{x_1,x_2,x_3} = \B \vec{e}_3
    \Theta(R - \sqrt{(x_1^2+x_2^2)}),
\end{equation}
where $\Theta$ is the Heaviside step function, $\vec{e}_3$ is the unit vector along the 3rd coordinate axis and $\B$ is the magnetic field strength. The Fourier transform of the magnetic field configuration (\ref{eq:B_solenoid}) can be calculated analytically:
\begin{equation}
\label{eq:B_k}
  \tilde{B}_3(\vec{k}) =\B R^2 \sqrt{\frac{\pi}{2}}\delta(k_3)
  \ _0F_1\lr{2;-\frac{1}{4}(k_1^2+k_2^2)R^2},
\end{equation}
where $_0F_1$ is a generalised hypergeometric function~\cite{OlverNIST}.

 According to the Kubo formula (\ref{cmc_Kubo}) and the expression (\ref{current_linear_response1}) (with an obvious permutation of indices), the Fourier transform of the 3rd component of the electric current is related to $\tilde{B}_3(\vec{k})$ as
\begin{equation}
\label{eq:J_k}
 \tilde{j}_3(\vec{k}) =  \scme\lr{k} \tilde{B}_3(\vec{k}) .
\end{equation}
Since here we are interested in a rather qualitative estimate of the electric current, for simplicity we use the continuum expression (\ref{eq:ft_current_current_PV}) for $\scme$, which describes our numerical data quite well in the phase with unbroken parity (see Fig.~\ref{fig:JJ_big}).

\begin{figure}[h!tpb]
  \centering
  \includegraphics[width=0.95\linewidth]{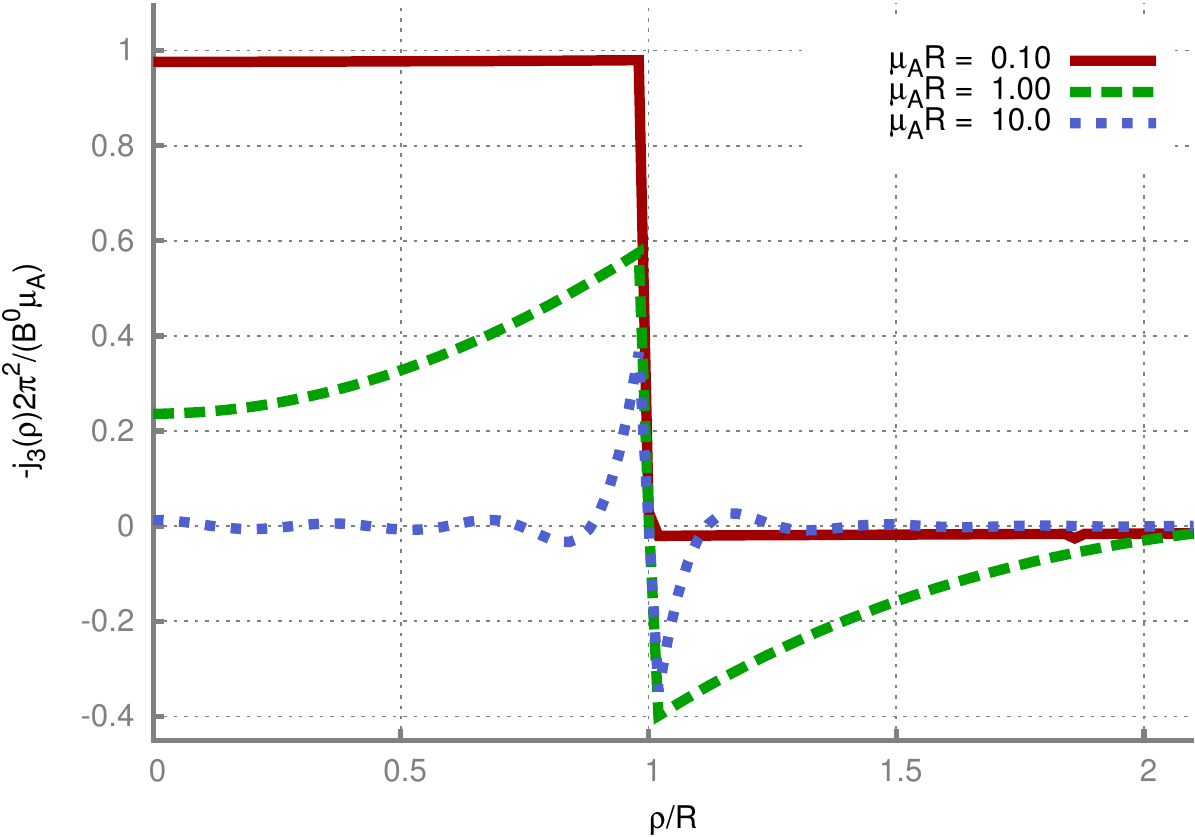}
  \caption{Electric current density $j_3\lr{\rho}$ in the direction of the magnetic field as a function of the radial coordinate $\rho$ for different values of the dimensionless quantity $\mua R$, where $R$ is the radius of the region with nonzero magnetic field. For convenience, we show the ratio of $j_3\lr{\rho}$ to the conventional value $j_3 = \frac{\mua B}{2 \pi^2}$ which follows from (\ref{cmc_def}) and (\ref{cmc_theory0}).}
  \label{fig:Jz_finitesize}
\end{figure}

 For experiments it is much more convenient to consider the coordinate space current $j_3(\vec{x})$. There is no analytic expression for the inverse Fourier transform of (\ref{eq:J_k}), and we have calculated it numerically. Because of the cylindrical symmetry of the magnetic field configuration the current $j_3(\vec{x}) = j_3(\rho)$ depends only on the radial coordinate $\rho = \sqrt{x_1^2+x_2^2}$ and on the parameters $R$, $\mua$ and $\B$. The magnetic field strength simply enters the final result as a factor, but $j_3(\rho)$ is a nontrivial function of $\mua$ and $R$. In Fig.~\ref{fig:Jz_finitesize} we plot $j_3(\rho)$ at fixed $\mua$ for different values of the dimensionless product $\mua R$. For small $\mua R$ the current inside the solenoid is almost constant and its magnitude is close to the value $\frac{\mua \B}{2 \pi^2}$ in complete agreement with the conventional expression (\ref{cmc_theory0}). As we increase $\mua R$ the current develops a stronger dependence on $\rho$ and assumes its maximum near the boundary of the solenoid. It is also interesting to note that the total current through the $\lr{x_1,x_2}$ plane is proportional to $\sigma_{CME}\lr{k \rightarrow 0}$ and hence vanishes. Thus the current inside the solenoid is compensated by a ``heavy-tailed'' current density of opposite sign outside of the solenoid.

 To find an estimate for $\mua$ in a typical experiment we can use the results of \cite{Kharzeev:14:1}, where a reasonably good description of experimental data was obtained under the assumption $\mua \ll \mu$ and the chemical potential was found to be $\mu \sim 100 \ \meV$. As a reasonable upper estimate for the chiral chemical potential we can take $\mua \sim 10 \meV$. To observe a strong chiral magnetic current inside the sample our numerical calculations suggest that one should have $\mua R \lesssim 1$.  To achieve this the sample size (in the directions orthogonal to the magnetic field) should be of order $R \sim 10^{2} \ \eV^{-1}$ or, in MKS units, $R \sim 10^{-2} \ \textrm{mm}$. This is a reasonable estimate if one takes into account the characteristic size of the sample of order $1 \textrm{mm}$ and possible inhomogeneities of crystal structure.

 Unfortunately, the characteristic values of the electric field strength in the sample are not reported in \cite{Kharzeev:14:1}, and hence it is difficult to estimate the stationary value of $\mua$ directly using the expression (\ref{muA_estimate}) derived from the anomaly equation. However, it is interesting to note that the above discussed dependence of the CME response on $\mu_A R$ should result in the nonlinear dependence of the electric current in the sample on the electric field strength. Indeed, for sufficiently small electric fields the stationary value of $\mua$ in (\ref{muA_estimate}) and hence the quantity $\mua R$ are small, and the CME response should be well described by the conventional expression (\ref{cmc_theory0}). With the increase of the electric field, however, the stationary value of $\mua$ and hence $\mua R$ increase, and according to Fig.~\ref{fig:Jz_finitesize} the CME contribution to the conductivity of the sample should decrease.

 In our simple estimates we have neglected several complications that can occur in realistic experiments. First of all, we have approximated the state of dynamical equilibrium in the chirality pumping process by a ground state of the many-body Hamiltonian, thus completely neglecting all dynamical real-time processes. We have also not taken into account that the magnetic field inside a sample depends on numerous different parameters and is in general a non-trivial function of the applied external field. Moreover we did not consider the effect of grain boundaries inside the sample or possible contributions of boundary states. Nevertheless our simple calculation shows that it is possible to observe the chiral magnetic effect even in the case of static and homogeneous magnetic fields and allows us to make educated order-of-magnitude estimates that could help in devising new experiments.

\section{Conclusions}
\label{sec:conclusions}

 In this paper we have presented the results of a mean field study of a lattice model for a parity-breaking Weyl semimetal with on-site inter-electron interactions, in which the chiral symmetry is only an emergent low-energy symmetry. We have studied the phase diagram of the model in the parameter space of bare mass, interaction strength and bare chiral chemical potential. Moreover, for all points of the phase diagram which we considered we have also calculated the static chiral magnetic conductivity within the linear response theory.

 We have found that our model exhibits a phase of spontaneously broken $\mathcal{C}\mathcal{P}$ symmetry (Aoki phase, Axionic insulator phase). The phase transition to this phase is of the second order and persists at finite bare chemical potential and finite temperature. Increasing the bare chiral chemical potential leads to a slight decrease in the critical interaction strength for the phase transition. Our calculations show a strong multiplicative renormalization of the chiral chemical potential in both phases. The slope of the renormalized chiral chemical potential changes at the phase boundary, but in both phases it strongly grows with the interaction strength. This is consistent with a previous study of a model with continuum Dirac fermions~\cite{Buividovich:14:1}.

 It turned out that in our model the enhancement of the chiral magnetic conductivity (as compared to the non-interacting case) is mostly caused by the growth of the renormalized chiral chemical potential $\mua$ with interaction strength. Our estimates of $\scme$ never exceed (by absolute value) the universal value $\frac{N_f \, \mua}{2 \pi^2}$ with the renormalized value of the chiral chemical potential and the appropriate number $N_f$ of Dirac cones in the energy spectrum. If we adhere to the quantum field-theoretical viewpoint on the renormalization, in which only the renormalized quantities are physically observable, we are tempted to conclude that inter-electron interactions in fact do not result in enhancement of the chiral magnetic conductivity. To some extent the renormalized value of $\mua$ can be indeed observable, e.g., in ARPES studies of the fermion dispersion relation. On the other hand, in condensed matter systems we can also have an independent way of estimating the bare value of $\mua$, e.g. by using the relation similar to (\ref{muA_estimate}) in the case of chirality pumping process, or from the strain-dependent parameters of the tight-binding model \cite{Balents:12:1}. It is therefore a particularly interesting question what is the counterpart of the multiplicative renormalization of $\mua$ in the situation when chirality imbalance is created dynamically, and whether it leads to some modification of the naive estimates of the CME contribution to the longitudinal electric conductivity based on (\ref{muA_estimate}) and (\ref{cmc_theory0}).

 Furthermore, we have found that spontaneous breaking of $\mathcal{C}\mathcal{P}$ symmetry in the axionic insulator phase (or Aoki phase in lattice QCD terminology) results in an immediate strong suppression of $\scme$. This conclusion is in sharp contrast to the behavior of $\scme$ found for continuum Dirac fermions with contact interactions in \cite{Buividovich:14:1}, where it was found that the CME response is particularly large in a phase with broken chiral symmetry. Thus it seems that the continuum chiral symmetry is important for such strong enhancement. It might be interesting therefore to consider the CME response also for lattice fermions with a continuum chiral symmetry, e.g. for staggered fermions \cite{Kogut:75:1}. However, in this case there are $N_f = 4$ fermionic species and only one $U\lr{1}$ symmetry, which invalidates the direct comparison with the analysis of \cite{Buividovich:14:1}. Yet another possibility would be to use overlap Hamiltonian of \cite{Creutz:01:1} with $N_f = 1$ and an exact $U\lr{1}$ chiral symmetry. However, from the point of view of condensed matter physics the non-local overlap Hamiltonian is a rather artificial construction. Moreover, chiral symmetry for such a Hamiltonian is generated by non-local transformations \cite{Creutz:01:1} and is therefore explicitly broken by local on-site interactions.

 We have also studied the importance of inter-fermion interactions and found that in those regions of the phase diagram where the size of the gap in the energy spectrum is small, the only practically important effect of interactions is the renormalization of the chiral chemical potential which enters the universal formula (\ref{cmc_theory0}). The relative magnitude of loop corrections does not exceed a few tenth of percent along the lines in the phase diagram where the gap size is zero.

 We have found that corrections to $\scme$ due to inter-electron interactions are only important when the gap size is quite large, and become comparable with the non-interacting result inside the axionic insulator phase with broken $\mathcal{C}\mathcal{P}$ symmetry (see Fig.~\ref{fig:JJ_comp_005}). Taking into account that spontaneous $\mathcal{C}\mathcal{P}$ symmetry breaking is in fact a remainder of spontaneous chiral symmetry breaking in lattice models with emergent low-energy chiral symmetry, this finding supports the theoretical expectation that the relation between $\scme$ and the axial anomaly coefficient might break upon chiral symmetry breaking \cite{Buividovich:14:1,Knecht:04:1}. However, it seems that this observation is not very relevant for practical experiments with Weyl/Dirac semimetals, since the CME response is very strongly suppressed exactly when the corrections to the chiral magnetic conductivity due to interactions become important.

 In addition, we have briefly considered the effects of finite temperature on the CME response. In general, we found that finite temperature tends to destroy the ``axion condensate'' in the Aoki phase and suppresses the chiral magnetic conductivity.

 This work should be considered as the rather crude attempt at the description of the recent experiments with Dirac semimetals \cite{Kim:13:1,Kharzeev:14:1,Xiong:13:1}, where chirality imbalance is created by chirality pumping in parallel electric and magnetic fields, and the CME response results in the negative magnetoresistance phenomenon. Since chirality pumping is a dynamical process, to describe such an experimental situation in a more systematical way, one should use the real-time linear response theory and adjust the mean-field approximation correspondingly. Moreover, one should take into account possible boundary effects, which are absent when using periodic boundary conditions. We leave these technically more advanced calculations for further work.

 Finally, let us discuss the assumption of spatially and temporally homogeneous Hubbard-Stratonovich fields that do not break rotational invariance, which we have made in our mean-field analysis. Some theoretical considerations \cite{Boyarsky:12:1,Yamamoto:13:1,Manuel:15:1} suggest that the system of chirally imbalanced fermions is unstable towards the decay of chirality imbalance at the expense of creating helical magnetic field configurations. In the static limit, which is justifiable at sufficiently late times after the decay, such spatially inhomogeneous configurations should be described by the Beltrami equation \cite{Sadofyev:13:1}. In such configurations, the CME current is expected to vanish \cite{Sadofyev:13:1}.

 Since fermionic condensates in (\ref{HubbardStratonovich2}) to some extent can imitate external gauge fields (if, e.g., ${\Phi_x \sim \vev{ \hat{\phi}^{\dag}_x \otimes \hat{\phi}_x } \sim \alpha_i A_i}$), one can expect that if such instability is indeed relevant, there can exist spatially inhomogeneous condensates with lower free energy than that of the homogeneous condensates considered in this work. Calculations in the holographic Sakai-Sugimoto model partially support the possibility of a non-homogeneous ground state in the presence of chiral chemical potential \cite{Kim:10:1,Zamaklar:11:1,Bayona:12:1}, although typically only for sufficiently large values of $\mua$ of order of the vector meson masses. To check for possible unstable directions in the space of all possible configurations of inhomogeneous condensates we have investigated the eigenvalues of the Hessian of the free energy $\frac{\partial^2 \mathcal{F}}{\partial \Phi_{y,B} \partial \Phi_{x,A}}$. We found no evidence for non-positive eigenvalues and conclude that the homogeneous field $\Phi^{\star}$ is at least a local minimum in the space of all possible fields. Moreover, in a recent work \cite{Buividovich:15:2} co-authored by one of us the real-time decay of chirality imbalance in the Hamiltonian similar to (\ref{hwdirac_manybody}) was considered, and no evidence for the formation of nontrivial spatially inhomogeneous gauge field configurations was found. Furthermore, in this work it was found that the decay of chirality imbalance is strongly suppressed for small values of the Fermi velocity. For all these reasons we believe that the inhomogeneous ground state is irrelevant for chirally imbalanced Dirac or Weyl semimetals.

\begin{acknowledgments}
 This work was supported by the S.~Kowalevskaja award from the Alexander von Humboldt-Stiftung. We would like to thank D.~Kharzeev and K.~Landsteiner for interesting and stimulating discussions.
\end{acknowledgments}

\appendix
\section{Second derivatives of the fermionic free energy over the gauge vector potential and the Hubbard-Stratonovich field}
\label{apdx:ffe_derivatives}

 In order to calculate the chiral magnetic conductivity $\scme\lr{k}$ from the Kubo formula (\ref{cmc_Kubo}), we need to calculate second derivatives of the free energy with respect to external fields according to the expression (\ref{ffe_2nd_der_general}). Since we are working with a finite lattice, we can perform summation over eigenstates in this expression numerically in a straightforward way.

 To this end we need an explicit expression for the effective single--particle Hamiltonian.
Substituting the saddle-point value of the  Hubbard-Stratonovich field (\ref{restr_saddle_point}) into the definition of the effective Hamiltonian (\ref{effective_sp_hamiltonian}), we obtain:
\begin{eqnarray}
\label{hwdirac_singlepart_apdx}
 h_{x, y}
 =
   \sum\limits_{k=1}^{3} -i \alpha_k \nabla_{k \, x y}
 +
   \frac{\gamma_0}{2} \Delta_{x y}
 + \nonumber \\ +
   \Phi_x \delta_{x,y}+
   \gamma_0 \mr + i \gamma_0 \gamma_5 \mi
 + \gamma_5 \mua,
\end{eqnarray}
and in the absence of external fields in momentum space:
\begin{eqnarray}
\label{hwdirac_singlepart_momentum_apdx}
 h\lr{k} = \sum\limits_{i=1}^{3} \alpha_i \sin\lr{k_i}
 +
 2 \, \gamma_0 \sum\limits_{i=1}^{3} \sin^2\lr{k_i/2} + \nonumber \\ +
 \gamma_0 \mr + i \gamma_0 \gamma_5 \mi
 + \gamma_5 \mua.
\end{eqnarray}
The energy levels of the Hamiltonian (\ref{hwdirac_singlepart_momentum_apdx}) are given by formula (\ref{eq:energy_levels}) and its eigenstates have the following form:
\begin{eqnarray}
\label{eigenstates}
\varphi_{s,\sigma}(\vec{k})=
\begin{pmatrix}
\sqrt{\frac{1}{2} + \frac{\sigma S - \mu_A}{2 \varepsilon_{s, \sigma}}} \eta_{\sigma} \\
s \frac{W+m_r - im_i}{\vert W+m_r + im_i\vert} \sqrt{\frac{1}{2} - \frac{\sigma S - \mu_A}{2 \varepsilon_{s, \sigma}}} \eta_{\sigma}
\end{pmatrix},
\end{eqnarray}
where $\eta_{\sigma}$ is the eigenstate of the operator $\sigma_i \mathrm{sin}(k_i)$ with eigenvalue $\sigma S$. The corresponding wavefunctions are given by:
\begin{eqnarray}
\Psi^{s,\sigma}_{x} (\vec{k}) = \varphi_{s,\sigma}(\vec{k})\frac{e^{i \vec{k} \cdot \vec{x}}}{\sqrt{L_s^3}} \label{eq:wdwavefuctions}.
\end{eqnarray}
To proceed further, we rewrite (\ref{ffe_2nd_der_general}) in more details:
\begin{eqnarray}
\label{ffe_2nd_der_det1}
 \frac{\partial^2 \mathcal{F}_0}{\partial \theta \, \partial \xi}=
 \sum\limits_{\vec{p},s,\sigma} n\lr{\varepsilon_{s,\sigma}(\vec{p})} \bra{s,\sigma,\vec{p}} \frac{\partial^2 h}{\partial \xi \, \partial \theta} \ket{s,\sigma,\vec{p}}
 - \nonumber \\ -
 \sum\limits_{\vec{p},s,\sigma} \frac{\bra{s,\sigma,\vec{p}} \frac{\partial h}{\partial \theta} \ket{s,\sigma,\vec{p}}
\bra{s,\sigma,\vec{p}} \frac{\partial h}{\partial \xi} \ket{s,\sigma,\vec{p}}}
{4 T \, \cosh^2\lr{\frac{\varepsilon_{s,\sigma}(\vec{p})}{2 T}}}
 +\nonumber \\+
\sum\limits_{\substack{\vec{p},s,\sigma \\ \vec{q},s^\prime,\sigma^\prime}} (n\lr{\varepsilon_{s,\sigma}(\vec{p})}-n\lr{\varepsilon_{s^\prime,\sigma^\prime}(\vec{q})})
 \times \nonumber \\ \times
 \frac{\bra{s,\sigma,\vec{p}} \frac{\partial h}{\partial \theta} \ket{s^\prime,\sigma^\prime,\vec{q}}
 \bra{s^\prime,\sigma^\prime,\vec{q}} \frac{\partial h}{\partial \xi} \ket{s,\sigma,\vec{p}}}{\varepsilon_{s,\sigma}(\vec{p}) - \varepsilon_{s^\prime,\sigma^\prime}(\vec{q})}
 ,
\end{eqnarray}
where the sum is taken over all momenta in the Brillouin zone and
\begin{eqnarray}
\label{matrix_elemnt_def}
\bra{s,\sigma,\vec{p}} \mathcal{O} \ket{s^\prime,\sigma^\prime,\vec{q}} =
\sum\limits_{x,y} \Psi^{\dagger, s,\sigma}_{x}(\vec{p}) \mathcal{O}_{x,y} \Psi^{s^{\prime},\sigma^{\prime}}_{y}(\vec{q}).
\end{eqnarray}

Now let us turn to the explicit form of the matrix elements in (\ref{ffe_2nd_der_det1}).
A simple calculation gives:
\begin{eqnarray}
\left. \frac{\partial h_{x,y}}{\partial A_{z,i}} \right|_{A=0}=
-(P_+\delta_{z+\hat{i}, x} \delta_{z, y}+
P_-\delta_{z, x} \delta_{z+\hat{i}, y}), \\
\left. \frac{\partial h_{x,y}}{\partial \Phi_{z,A}} \right|_{\Phi=0}= \delta_{z, x} \delta_{z,y} \Gamma_A,
\end{eqnarray}
and
\begin{eqnarray}
\left. \frac{\partial^2 h_{x,y}}{\partial A_{z_1,i} \partial A_{z_2,j}} \right|_{A=0}= \nonumber \\ =
i \delta_{z_1, z_2} \delta_{i,j} \left(
P_-\delta_{z_1, x} \delta_{z_1+\hat{i}, y}-
P_+\delta_{z_1+\hat{i}, x} \delta_{z_1, y}
\right),\\
\frac{\partial^2 h_{x,y}}{\partial A_{z_1,i} \partial \Phi_{z_2,A}} = \frac{\partial^2 h_{x,y}}{\partial \Phi_{z_1,A} \partial A_{z_2,i}}=0,\\
\frac{\partial^2 h_{x,y}}{\partial \Phi_{z_1,A} \partial \Phi_{z_2,B}}=0,
\end{eqnarray}
where we denote
\begin{eqnarray}
P_{\pm} = \frac{\alpha_i \pm i \gamma_0}{2}.
\end{eqnarray}
Using expression (\ref{matrix_elemnt_def}), we can represent the matrix elements in (\ref{ffe_2nd_der_det1}) as:
\begin{eqnarray}
\bra{s,\sigma,\vec{p}}\frac{\partial h}{\partial A_{z,i}} \ket{s^\prime,\sigma^\prime,\vec{q}} = \nonumber \\ =
\bar{\varphi}_{s, \sigma}(\vec{p})
j_i
\varphi_{s^{\prime}, \sigma^{\prime}}(\vec{q})
\frac{e^{-i(\vec{p} - \vec{q})\cdot \vec{z}}}{L_s^3},\label{eq:dh_dA}\\
\bra{s,\sigma,\vec{p}}\frac{\partial h}{\partial \Phi_{z,A}} \ket{s^\prime,\sigma^\prime,\vec{q}} = \nonumber \\ =
\bar{\varphi}_{s, \sigma}(\vec{p})\Gamma_A \varphi_{s^{\prime}, \sigma^{\prime}}(\vec{q}) \frac{e^{-i(\vec{p} - \vec{q})\cdot \vec{z}}}{L_s^3},\\
\bra{s,\sigma,\vec{p}}\frac{\partial^2 h}{\partial A_{z_1,i} \partial A_{z_2,j}} \ket{s,\sigma,\vec{p}} = \nonumber \\ =
\bar{\varphi}_{s, \sigma}(\vec{p})
\frac{\partial j_i}{\partial A_j}
\varphi_{s, \sigma}(\vec{p})\frac{\delta_{z_1,z_2}}{L_s^3} \label{eq:d2h_dAdA},
\end{eqnarray}
where
\begin{eqnarray}
j_i = -(P_+ e^{-i \vec{p} \cdot \vec{e}_i} + P_- e^{i \vec{q} \cdot \vec{e}_i}), \\
\frac{\partial j_i}{\partial A_j} = i \delta_{i,j} (P_-  e^{i \vec{p} \cdot \vec{e}_i} - P_+ e^{-i \vec{p} \cdot \vec{e}_i}),
\end{eqnarray}
and $\vec{e}_i$ is a unit lattice vector in the direction $i$.

Considering for instance the derivative $\partial^2 \mathcal{F}_0 / \partial A_i \partial A_j$, we substitute expressions (\ref{eq:dh_dA}) and (\ref{eq:d2h_dAdA}) into (\ref{ffe_2nd_der_det1}) and perform a Fourier transform with respect to $z_1$ and $z_2$:
\begin{eqnarray}
\frac{\partial^2 \mathcal{F}_0}{\partial A_i \, \partial A_j} (\vec{k})
 = \nonumber \\ =
\frac{1}{L_s^3}\sum\limits_{z_1,z_2}e^{i\vec{k}(\vec{z_1}-\vec{z_2})} \frac{\partial^2 \mathcal{F}_0}{\partial A_i \, \partial A_j} (z_1, z_2),
\end{eqnarray}
and obtain the final expression, which can be easily calculated numerically:
\begin{eqnarray}
\label{eq:KuboFormula}
L_s^3 \frac{\partial^2 \mathcal{F}_0}{\partial A_i \, \partial A_j} (\vec{k})=\nonumber \\ =
\sum\limits_{s,\sigma}\sum\limits_{\vec{p}} \Biggl[
 \bar{\varphi}_{s, \sigma}(\vec{p}) \frac{\partial j_i}{\partial A_j}  \varphi_{s, \sigma}(\vec{p})
 n\lr{\varepsilon_{s,\sigma}(\vec{p})}
- \nonumber \\ -
\delta_{\vec{k},0} \frac{\bar{\varphi}_{s, \sigma}(\vec{p}) j_i  \varphi_{s, \sigma}(\vec{p})
 \bar{\varphi}_{s, \sigma}(\vec{p}) j_j  \varphi_{s, \sigma}(\vec{p})}
{4 T \, \cosh^2\lr{\frac{\varepsilon_{s,\sigma}(\vec{p})}{2 T}}}
 + \nonumber \\ +
 \sum\limits_{s^{\prime}, \sigma^{\prime}}
 \frac{\bar{\varphi}_{s, \sigma}(\vec{p}) j_i  \varphi_{s^{\prime}, \sigma^{\prime}}(\vec{q})
 \bar{\varphi}_{s^{\prime}, \sigma^{\prime}}(\vec{q}) j_j  \varphi_{s, \sigma}(\vec{p})}
 {\varepsilon_{s,\sigma}(\vec{p}) - \varepsilon_{s^{\prime},\sigma^{\prime}}(\vec{q})} \times \nonumber \\ \times
 \left( n\lr{\varepsilon_{s,\sigma}(\vec{p})} - n\lr{\varepsilon_{s^{\prime},\sigma^{\prime}}(\vec{q})}\right) \Biggr],
\end{eqnarray}
where $\vec{q} = \vec{p} + \vec{k}$ and the sum over momentum $\vec{p}$ is taken over the Brillouin zone.

Final expressions for the derivatives $\partial^2 \mathcal{F}_0 / \partial A_i \partial \Phi_A$ and $\partial^2 \mathcal{F}_0 / \partial \Phi_A \partial \Phi_B$ can be derived in the same manner.

\section{The role of the Fermi velocity}
\label{apdx:fermi_velocity}

 In this Appendix we demonstrate that the Fermi velocity $v_F$ in the Hamiltonian (\ref{hwdirac_manybody}) amounts to a simple rescaling of the model parameters and observables, which is a general feature of the instantaneous potential approximation for inter-fermion interactions. Therefore in our calculations we use $v_F  = 1$. In this Appendix we provide explicit expressions which can be used to restore the dependence on $v_F$ in all our results.

 It is easy to check that if we substitute the rescaled values
\begin{eqnarray}
\label{rescaling1}
 \bar{m}^{\lr{0}}    = m^{\lr{0}}/v_F, \quad
 \bar{\mu}_A^{\lr{0}} = \mua^{\lr{0}}/v_F, \quad
 \bar{U}             = U/v_F
\end{eqnarray}
into the Hamiltonian (\ref{hwdirac_manybody}) with the unit Fermi velocity $v_F = \bar{v}_F \equiv 1$, we obtain the same Hamiltonian with a non-unit value of $v_F$, but rescaled by $1/v_F$:
\begin{eqnarray}
\label{rescaling2}
 \hat{H}\lr{v_F=1, \bar{m}^{\lr{0}}, \bar{\mu}_A^{\lr{0}}, \bar{U} }
 = \nonumber \\ =
 \frac{1}{v_F} \hat{H}\lr{v_F, m^{\lr{0}}, \mua^{\lr{0}}, U} .
\end{eqnarray}
It is also easy to check that the same rescaling works also upon the Hubbard-Stratonovich transformation, if we rescale the Hubbard-Stratonovich field as $\bar{\Phi}_{x} = \Phi_x/v_F$. In particular this implies that if the renormalized model parameters $\bar{\mu}_A$, $\bar{m}_r$ and $\bar{m}_i$ are known at $v_F = \bar{v}_F \equiv 1$, their values at non-unit Fermi velocity can be found simply as
\begin{eqnarray}
\label{rescaling_renormalized}
 m_r   = v_F \bar{m}_r, \quad
 m_i   = v_F \bar{m}_i, \quad,
 \mu_A = v_F \bar{\mu}_A.
\end{eqnarray}

 Substituting the scaling law (\ref{rescaling2}) into the definition $\mathcal{F} = -T \ln\lr{\tr \expa{-\hat{H}/T}}$ of the free energy, it is also easy to obtain
\begin{eqnarray}
 \mathcal{F}\lr{v_F, m^{\lr{0}}, \mua^{\lr{0}}, U, T}
 = \nonumber \\ =
 v_F \mathcal{F}\lr{v_F=1, \bar{m}^{\lr{0}}, \bar{\mu}_A^{\lr{0}}, \bar{U}, \bar{T}} ,
\end{eqnarray}
where $\bar{T} = T/v_F$ is the rescaled temperature. Current-current correlators which enter the Kubo formula (\ref{cmc_Kubo}) are obtained as variations of the free energy over the external vector potential $\vec{A}\lr{x}$, which is replaced by the link phases $A_{x,k}$ upon Peierls substitution. From the explicit form of the Wilson-Dirac Hamiltonian (\ref{hwdirac_singlepart}) it is obvious that $A_{x,k}$ should not be rescaled when the Fermi velocity $v_F$ is replaced by $\bar{v}_F = 1$. We then immediately obtain
\begin{eqnarray}
\label{rescaling3}
 \vev{j_{x,k} j_{y,l}}\lr{v_F, m^{\lr{0}}, \mua^{\lr{0}}, U, T}
 = \nonumber \\ =
 \frac{\delta^2 \mathcal{F}\lr{v_F, m^{\lr{0}}, \mua^{\lr{0}}, U, T}}{\delta A_{x, k} \delta A_{y,l}}
 = \nonumber \\ =
 v_F \, \frac{\delta^2 \mathcal{F}\lr{v_F=1, \bar{m}^{\lr{0}}, \bar{\mu}_A^{\lr{0}}, \bar{U}, \bar{T}}}{\delta A_{x, k} \delta A_{y,l}}
 = \nonumber \\ =
 v_F \vev{j_{x,k} j_{y,l}}\lr{v_F=1, \bar{m}^{\lr{0}}, \bar{\mu}_A^{\lr{0}}, \bar{U}, \bar{T}} .
\end{eqnarray}
Since the coordinates $x$ and hence also the wave vector $k$ in the Kubo formula (\ref{cmc_Kubo}) do not scale with Fermi velocity, the equation (\ref{rescaling3}) immediately translates into
\begin{eqnarray}
\label{cmc_rescaling}
 \scme\lr{k, v_F, m^{\lr{0}}, \mua^{\lr{0}}, U, T}
 = \nonumber \\ =
 v_F \, \scme\lr{k, v_F=1, \bar{m}^{\lr{0}}, \bar{\mu}_A^{\lr{0}}, \bar{U}, \bar{T}} .
\end{eqnarray}

  Using (\ref{cmc_rescaling}) and (\ref{rescaling_renormalized}), (\ref{rescaling1}) one can now easily see that the linear term in the expansion of $\scme\lr{k, v_F, m^{\lr{0}}, \mua^{\lr{0}}, U, T}$ in powers of either the bare or the renormalized chiral chemical potential does not depend on Fermi velocity. It could be expected, since the derivative of $\scme$ over $\mu_A$ is related to the anomaly coefficient which also does not depend on the Fermi velocity.

\section{The fate of the Aoki fingers}
\label{apdx:aoki}

 For the sake of completeness we discuss the finite volume dependence of the ``Aoki fingers'' for $\mub = 0$. In the paper \cite{Aoki:84:1} Aoki presents a conjecture for a phase diagram of lattice QCD with Wilson fermions. Based on calculations in the two-dimensional Gross-Neveu model and an effective model of lattice QCD he suggests a phase diagram with two phases. In the notation of our paper the order parameter for the phase transition is the $\mathcal{C}\mathcal{P}$ breaking mass term $\mi$, which is zero in one phase and assumes a finite value in the other phase (Aoki phase, Axionic insulator phase). We work with the Hamiltonian formalism, where time is continuous and not discretized. In the  conjectured phase diagram in $D$ dimensions the Aoki phase then forms $D$ fingers that touch the bare mass axis on characteristic points as the coupling $U$ goes to zero.

We now give a heuristic argument why, in general,  the fingers do not extend all the way down to touch the bare mass axis. At the border of the Aoki phase the pion mass, which is given by $\mP^2= \frac{\partial^2\mathcal{F}}{\partial \mi^2}$, has to vanish. In $D=d+1$ dimensions the second derivative of the mean-field free energy (\ref{eq:mean_field_free_energy}) reads
\begin{eqnarray}
  \label{eq:deriv_free_energy}
   \frac{\partial^2\mathcal{F}}{\partial \mi^2} = \frac{1}{\ls^d}\frac{1}{2}\sum\limits_{\vec{k},\sigma}\left(\frac{1}{\varepsilon_{-1,\sigma}(\vec{k})}-\frac{\mi^2}{\varepsilon_{-1,\sigma}(\vec{k})^3}\right)-\frac{2}{U},
\end{eqnarray}
where $\vec{k}$ is a $d$-dimensional vector and we sum over all $\vec{k}$ in the Brilloin zone.
Let us now consider the limit $\ls \to \infty$. The sum in (\ref{eq:deriv_free_energy}) then becomes an
integral:
\begin{eqnarray}
  \label{eq:sum_to_int}
  \frac{1}{\ls^d}\sum\limits_{\vec{k}} \to \frac{1}{(2\pi)^d} \int d\vec{k} \propto \int dk \, k^{(d-1)}
\end{eqnarray}
In the case of vanishing $\mub$ the energy levels become degenerate and we define $\varepsilon_{-1}(\vec{k}) := \varepsilon_{-1,+1}(\vec{k}) = \varepsilon_{-1,-1}(\vec{k})$. Near the phase boundary $\mi \ll 1$ and around points where $W+\mr = 0$ the integral is dominated by the contributions from $k \ll 1$ and we can write
\begin{eqnarray}
  \label{eq:approx_free_energy}
  \frac{\partial^2\mathcal{F}}{\partial \mi^2}
  \approx \frac{2}{U}
  - \nonumber \\ -
  \gamma \int
  dk \, k^{(d-1)}\left(\frac{1}{\sqrt{k^2+\mi^2}} -
    \frac{\mi^2}{\sqrt{(k^2+\mi^2)^3}}\right) \nonumber,
\end{eqnarray}
where we collect the factors of $1/(2 \pi)$ and the contribution from the solid angle integral in the (positive) constant $\gamma$. The exact numerical value of $\gamma$ is irrelevant for the following argument. 

\begin{figure}[h!tpb]
  \centering
  \includegraphics[width=0.9\linewidth]{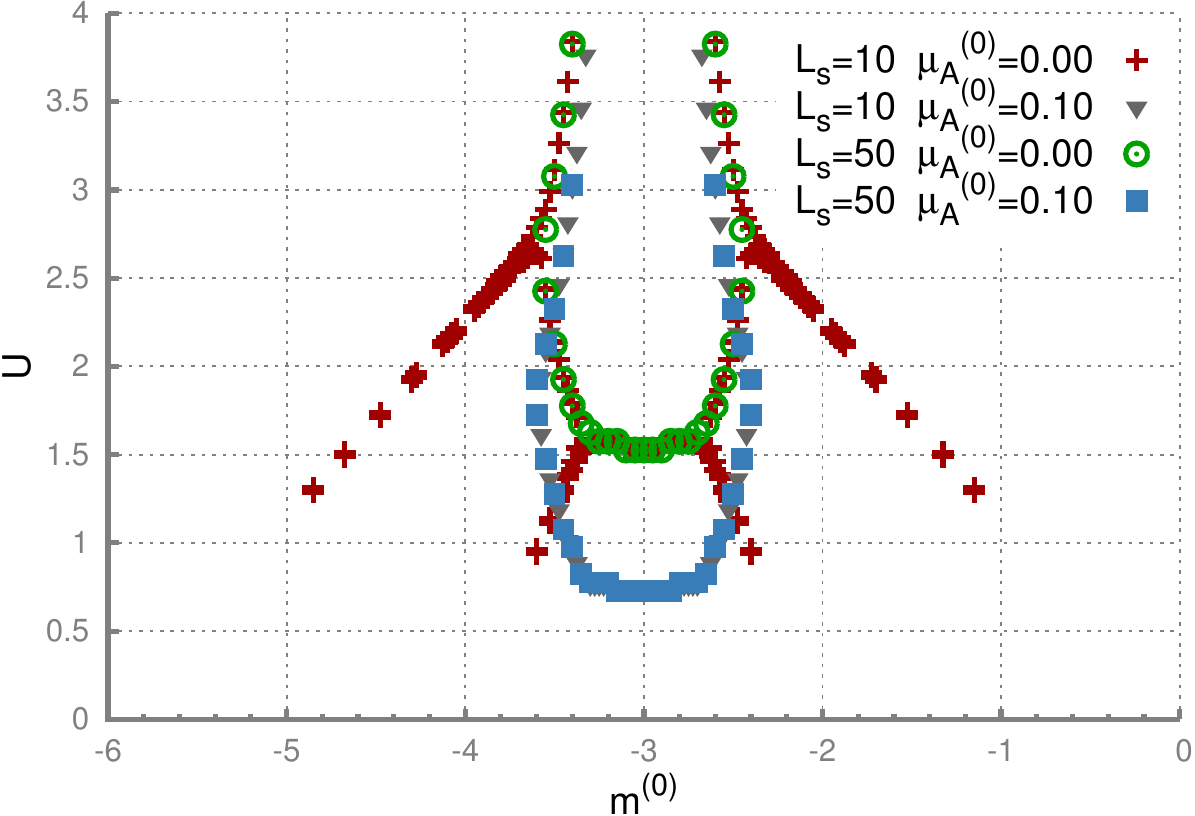}
  \caption{Volume dependence of the mean-field phase diagram. The Aoki fingers are volume dependent and seem to vanish as $\ls \to \infty$. This is only relevant if $\mub=0$. At finite $\mub$ the fingers are not present and phase diagrams for different lattice volumes lie on top of each other. In the figure the phase boundaries for $\mub=0.10$ where shifted by $-0.8$ in $U$ for better visibility.}
  \label{fig:phdiag}
\end{figure}

 As we approach the bare mass axis the term $2/U$ diverges. In order to get $\partial^2 \mathcal{F} / \partial \mi^2 = \mP^2 = 0$ this divergence has to be canceled by the integral. In $D = 1 + 1$ dimensions the first term diverges like $\log (\mi)$ and the second term becomes simply a constant as  $\mi \to 0$. It is possible to cancel the divergence and the Aoki fingers can reach all the way down to the bare mass axis. We performed numerical calculations in $D = 1+1$ dimensions in the limit $\ls \to \infty$ and found that the results are consistent with that statement that the fingers touch the bare mass axis.

In $D = 3 + 1$ dimensions, however, the first integral is no longer divergent as $\mi \to 0$. The contribution from the second integral is always non-negative and can not cancel the divergence. It is therefore not possible to find roots of equation (\ref{eq:deriv_free_energy}) and the Aoki phase can not extend down to the bare mass axis.

With the same argument it is now straight forward to see why the Aoki fingers are volume dependent. If $\ls$ is small the sum in (\ref{eq:deriv_free_energy}) diverges as $1/\sqrt{k^2}$ for vanishing $\mi$ and we can find a solution for $\partial^2 \mathcal{F} / \partial \mi^2  = 0$ even for small $U$. As we increase $\ls$ the sum becomes a better and better approximation of a (non-diverging) integral and (\ref{eq:deriv_free_energy}) does no longer have roots for $U \ll 1$.

%\bibliography{Buividovich}
%\bibliographystyle{apsrev}

\end{document}